\documentclass[useAMS,usenatbib]{mnras}

\usepackage{amssymb}
\usepackage{amsmath}
\usepackage{amsfonts}
\usepackage{graphicx}
\usepackage{caption}
\usepackage{subfig}
\usepackage{color}
\usepackage{multirow}

\def\apj{ApJ}
\def\apjl{ApJ}
\def\apjs{ApJS}
\def\aap{A\&A}
\def\mnras{MNRAS}
\def\nat{Nature}

\def\FeH{\mathrm{[Fe/H]}}
\def\aFe{[\alpha/\mathrm{Fe}]}

\DeclareMathOperator{\sech}{sech}

\newcommand{\kms}{\,{\rm km\, s^{-1}}}

\newcommand{\kpc} {\,\mathrm{kpc}}

{\newif\ifnotend
\notendtrue
\def\veclist{ABCDEFGHIJKLMNOPQRSTUVWXYZabcdefghijklmnopqrstuvwxyz.}
\def\top#1#2.{#1}
\def\tail#1#2.{#2.}
\loop\expandafter\xdef\csname v\expandafter\top\veclist\endcsname%
{{\noexpand\bf\expandafter\top\veclist}}
\edef\veclist{\expandafter\tail\veclist}
\if\veclist.\notendfalse\fi\ifnotend\repeat}

\title[The first pericentric passage of Sagittarius]{The outer low-$\alpha$ disc of the Milky Way - I: evidence for the first pericentric passage of Sagittarius?}

\author[P. Das et al. ]{Payel Das$^{1},$\thanks{E-mail:p.das@surrey.ac.uk}
Yang Huang$^{2,3}$,
Ioana Ciuc\u{a}$^{4,5,6}$ and Francesca Fragkoudi$^{7,8}$
\\
$^{1}$Astrophysics Research Group, University of Surrey, Guildford, Surrey, GU2 7XH\\
$^{2}$School of Astronomy and Space Science, University of Chinese Academy of Sciences, Beijing 100049, P.\,R.\,China\\
$^{3}$Key Lab of Optical Astronomy, National Astronomical Observatories, Chinese Academy of Sciences, Beijing 100012, P.\,R.\,China\\
$^{4}$Research School of Astronomy \& Astrophysics, Australian National University, Canberra, ACT 2611, Australia \\
$^{5}$School of Computing, Australian National University, Canberra, ACT 2601, Australia \\
$^{6}$ARC Centre of Excellence for All Sky Astrophysics in 3 Dimensions (ASTRO 3D), Australia \\
$^{7}$Institute for Computational Cosmology, Durham University, South Road, Durham DH1 3LE, UK \\
$^{8}$Department of Physics, Durham University, South Road, Durham DH1 3LE, UK \\
}

\begin{document}

\pagerange{\pageref{firstpage}--\pageref{lastpage}} \pubyear{2023}

\maketitle

\label{firstpage}

\begin{abstract}
Phase-space data, chemistry, and ages together reveal a complex structure in the outer low-$\alpha$ disc of the Milky Way. The age-vertical velocity dispersion profiles beyond the Solar Neighbourhood show a significant jump at $6\,\text{Gyr}$ for stars beyond the Galactic plane. Stars older than $6\,\text{Gyr}$ are significantly hotter than younger stars. The chemistry and age histograms reveal a bump at $\FeH = -0.5$, $\aFe = 0.1$, and an age of $7.2\,\text{Gyr}$ in the outer disc. Finally, viewing the stars beyond $13.5\,\text{kpc}$ in the age-metallicity plane reveals a faint streak just below this bump, towards lower metallicities at the same age. Given the uncertainty in age, we believe these features are linked and suggest a pericentric passage of a massive satellite $\sim 6\,\text{Gyr}$ ago that heated pre-existing stars, led to a starburst in existing gas. New stars also formed from the metal-poorer infalling gas. The impulse approximation was used to characterise the interaction with a satellite, finding a mass of $\sim 10^{11} M_{\odot}$, and a pericentric position between $12$ and $16\,\text{kpc}$. The evidence points to an interaction with the Sagittarius dwarf galaxy, likely its first pericentric passage.
\end{abstract}


\section[]{Introduction} 

The third data release of Gaia has given us access to parallaxes and proper motions for a billion stars \citep{gaia+16,arenou+18,gaia+18,lindegren+18,gaia+21}. Millions of these stars have complementary spectroscopic data from a range of surveys, for which ages and distances have been estimated using isochrones \citep[e.g.][]{mints18,quieroz+18,sanders+18,kordopatis+23} and ages have been estimated for red giant samples using data-driven spectroscopic estimators \citep{das+19,huang+20,ciuca+21}. Six-dimensional phase-space coordinates, metallicity, a number of abundances, and age have thus become available for an extraordinary number of stars, enabling us to characterize the chemodynamical behaviour of stars in detail both within the Solar Neighbourhood and beyond. The outer disc of the Milky Way has been revealed to unprecedented detail for the first time.

The low-$\aFe$ stars in our Milky Way have been shown to flare, i.e. the vertical scale height of the stars increases with radius \citep[e.g.][]{bovy+16,thomas+19,zheng+21}. The scale height varies by a factor of two between the Solar Neighbourhood and a Galactocentric distance of $12\,\text{kpc}$ \citep{amores+17}. It exists across all ages and is more pronounced for the metal-poorer ($\FeH\sim-0.6$) stars \citep{mackereth+17}. The radial velocity dispersion profile falls exponentially throughout for the oldest stars, while the youngest stars plateau beyond the solar radius \citep{sanders+18,sharma+21}. The vertical velocity dispersion profile falls exponentially until just beyond the solar radius and then plateaus for the oldest stars and even rises again for the youngest stars. This is partially attributable to a selection function effect, as stars at a larger heliocentric distance will be observed at higher vertical heights, and are therefore likely to have higher vertical velocity dispersions. However, one may also expect a flat vertical velocity dispersion, under the assumption of approximate equilibrium. If we imagine the vertical structure of the outer disc to be approximately described by vertical slab dynamics, then the vertical velocity dispersion is proportional to the dependence of the vertical height on Galactocentric radius, multiplied by the slope of the circular velocity curve. The vertical height increases roughly as $R$, and the circular velocity curve is dominated by dark matter, which decreases roughly as $1/R$ for an NFW profile. Therefore the vertical velocity dispersion profile should be approximately constant.

The flare in the outer low-$\aFe$ disc could be a consequence of the radial migration of stars from the inner disc. \cite{minchev+12a} investigated the evolution of galactic discs in N-body simulations. They found that the outskirts of discs comprised of stars migrated from the inner disc would have relatively large radial velocity dispersions. However, other work \citep{minchev+12b,veraciro+14,veraciro+16} found that inward migrators thin down as they move in, whereas outward migrators preserve the disc scale height at their destination. 

Flaring has also been shown in a number of N-body simulations to result from interactions between the disc and satellites \citep{velazquez+99,kazantzidis+09,gomez+13}. \cite{martig+14a} studied seven simulated disc galaxies, three with a quiescent merger history, and four with mergers in the last $9\,\text{Gyr}$. All galaxies undergo a phase of merging activity at high redshift, so that stars older than $9\,\text{Gyr}$ are found in a centrally concentrated component, while younger stars are mostly found in discs. They find that the lack of flaring did not necessarily imply a merger-free history, but mergers produce discontinuities between thin and thick disc components, and jumps in the age-velocity relation \citep{martig+14b}. 

\cite{church+19} compared the heating effect of ultralight axion-like particles or `fuzzy dark matter (FDM)' to cold dark matter (CDM) sub-haloes. They found that both produce a flaring that terminates the Milky Way disc at $15-20 \,\text{kpc}$. \cite{chiang+23} present a more sophisticated treatment that incorporates the full baryon and dark matter distributions of the Milky Way, and adopts stellar disc kinematics inferred from recent Gaia, APOGEE, and LAMOST surveys. They found that repeated sub-halo passages drives flaring of the outer disc and can even result in the observed `U-shaped' disc vertical velocity dispersion profile \citep{sanders+18}.  

\cite{renaud+21b} find in \texttt{VINTERGATAN}, a cosmological simulation of a Milky-Way-like galaxy, that the outer gas disc is fuelled by a distinct metal-poorer gas flow to the inner disc. The first passage of the last major merger galaxy triggers star formation. These in-situ stars have halo-like kinematics, due to the inclination of the outer disc that eventually aligns with the inner one via gravitational torques, therefore simulating a flared disc.

Here, we use a sample of $\sim140,000$ primary red clump (RC) stars drawn from the  Large Sky Area Multi-Object fiber Spectroscopic Telescope (LAMOST) survey \citep{huang+20}, which have precise distances and proper motions from Gaia DR3 and machine-learned ages. This group of stars uniquely comprises primarily flared thin-disc stars, as they are metal rich and belong to the Galactic anti-centre. We select low-$\alpha$ stars, and explore their phase-space, chemical and age structure in Section \ref{sec:struct}. We calculate robust vertical velocity dispersion profiles in Section \ref{sec:sigmaz}. We find evidence for a jump in the vertical velocity dispersion at $6\,\text{Gyr}$, for stars in the outer disc and beyond the Galactic plane. We attribute this to an interaction with a massive satellite, and characterize its position and mass using the impulse approximation in Section \ref{sec:satellite_interaction}. We discuss our results in the context of relevant literature in Section \ref{sec:discussion}, and conclude in Section \ref{sec:conc}. 

\begin{figure}
    \centering
    \includegraphics[width=1.0\columnwidth]{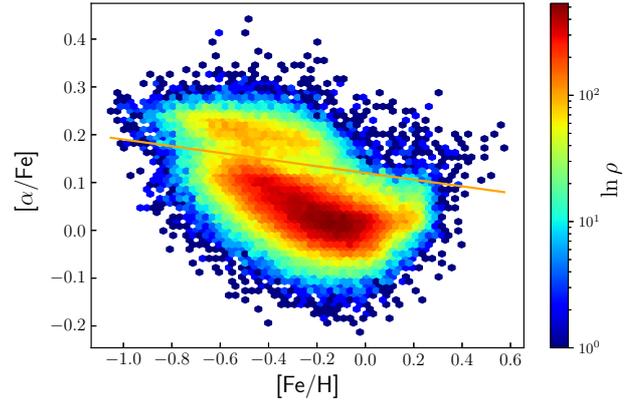}
    \caption{A log-density map of the primary RC sample in the $\aFe$-$\FeH$ plane. The orange line divides the low- and high-$\alpha$ populations of stars. The low-$\alpha$ primary RC sample contains $108,106$ stars.}\label{fig:afefehplane}
\end{figure}
\begin{figure*}
    \centering
    \subfloat[]{\includegraphics[width=1.0\columnwidth]{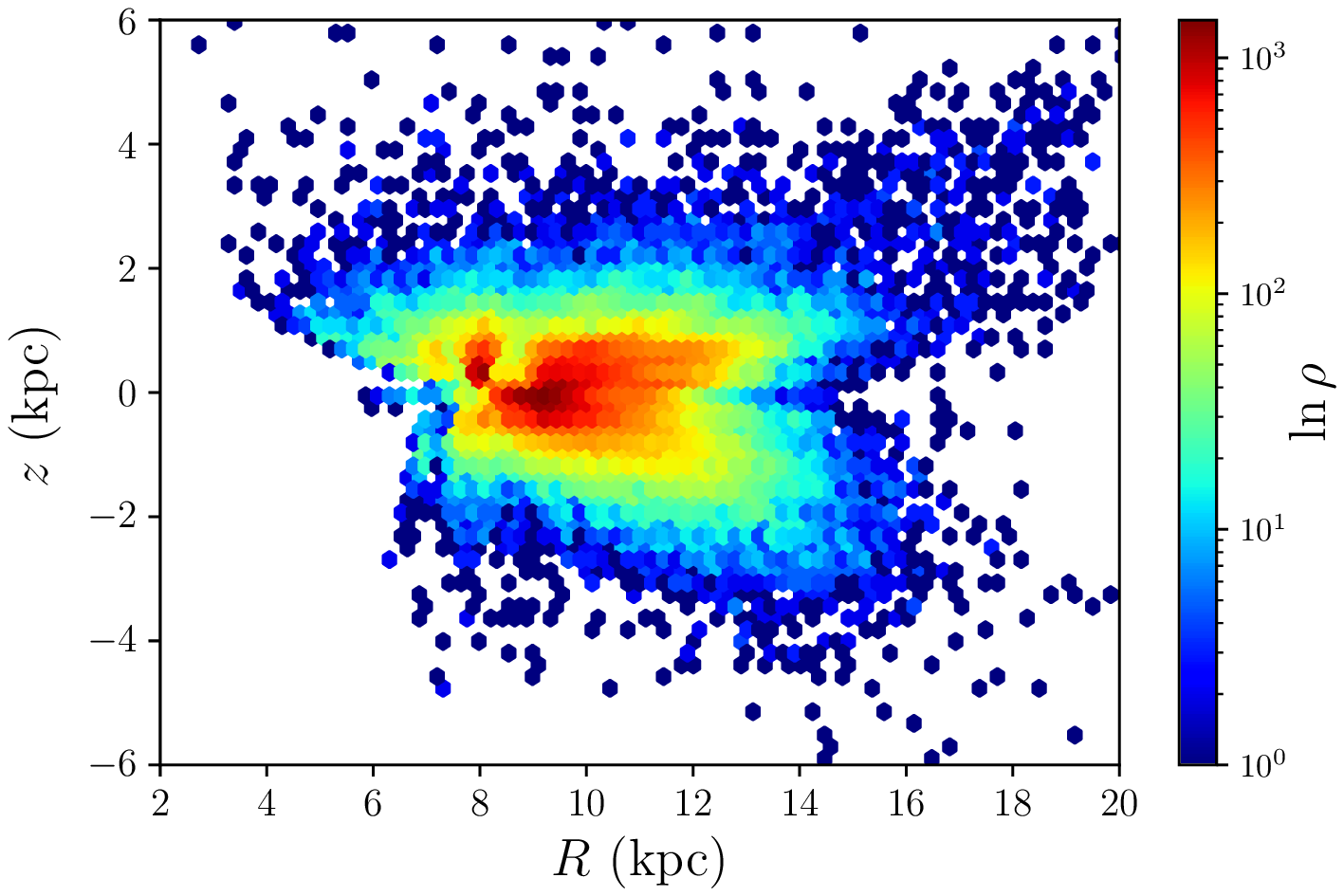}}
    \subfloat[]{\includegraphics[width=1.0\columnwidth]{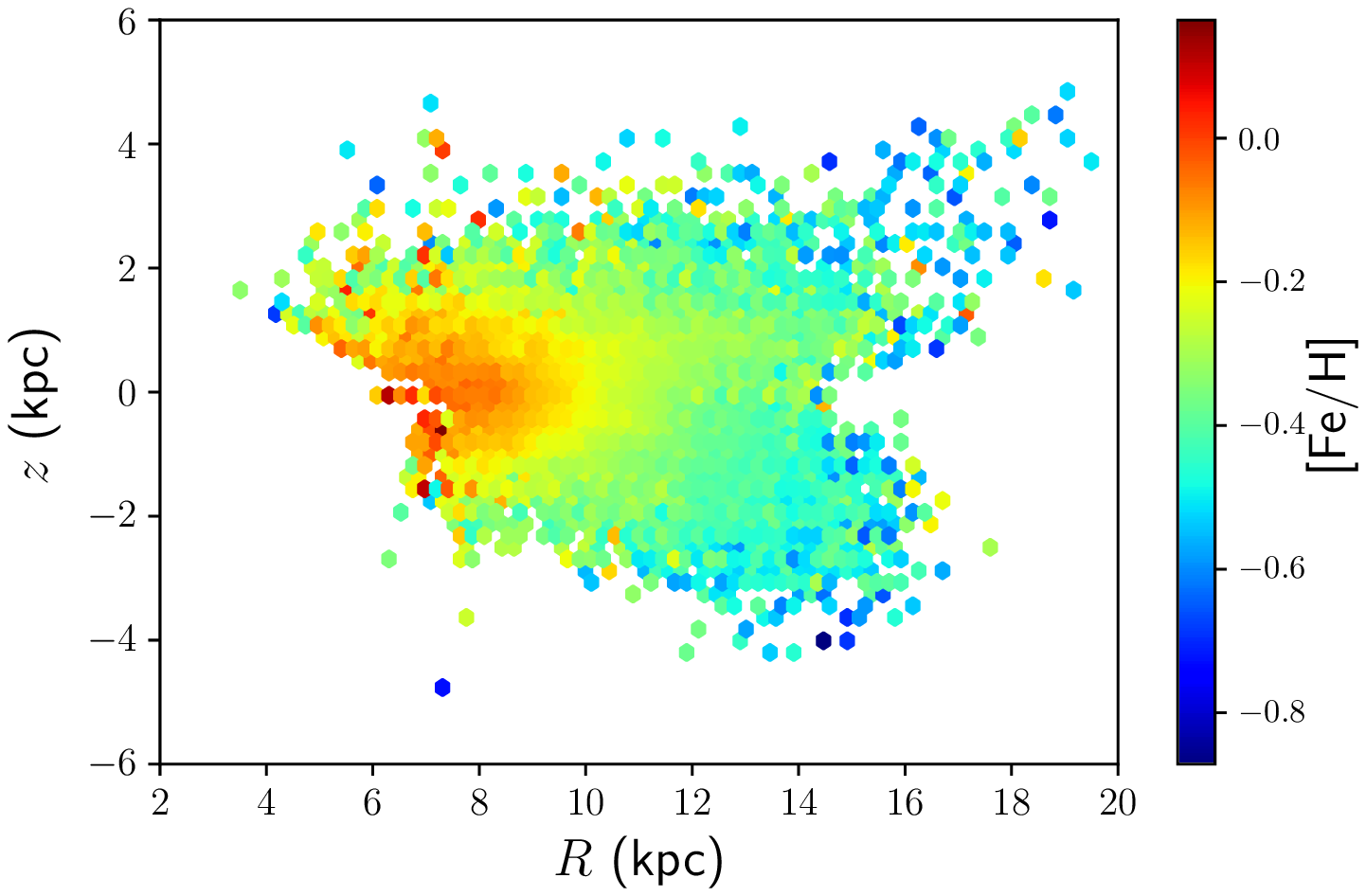}}
    \newline
    \subfloat[]{\includegraphics[width=1.0\columnwidth]{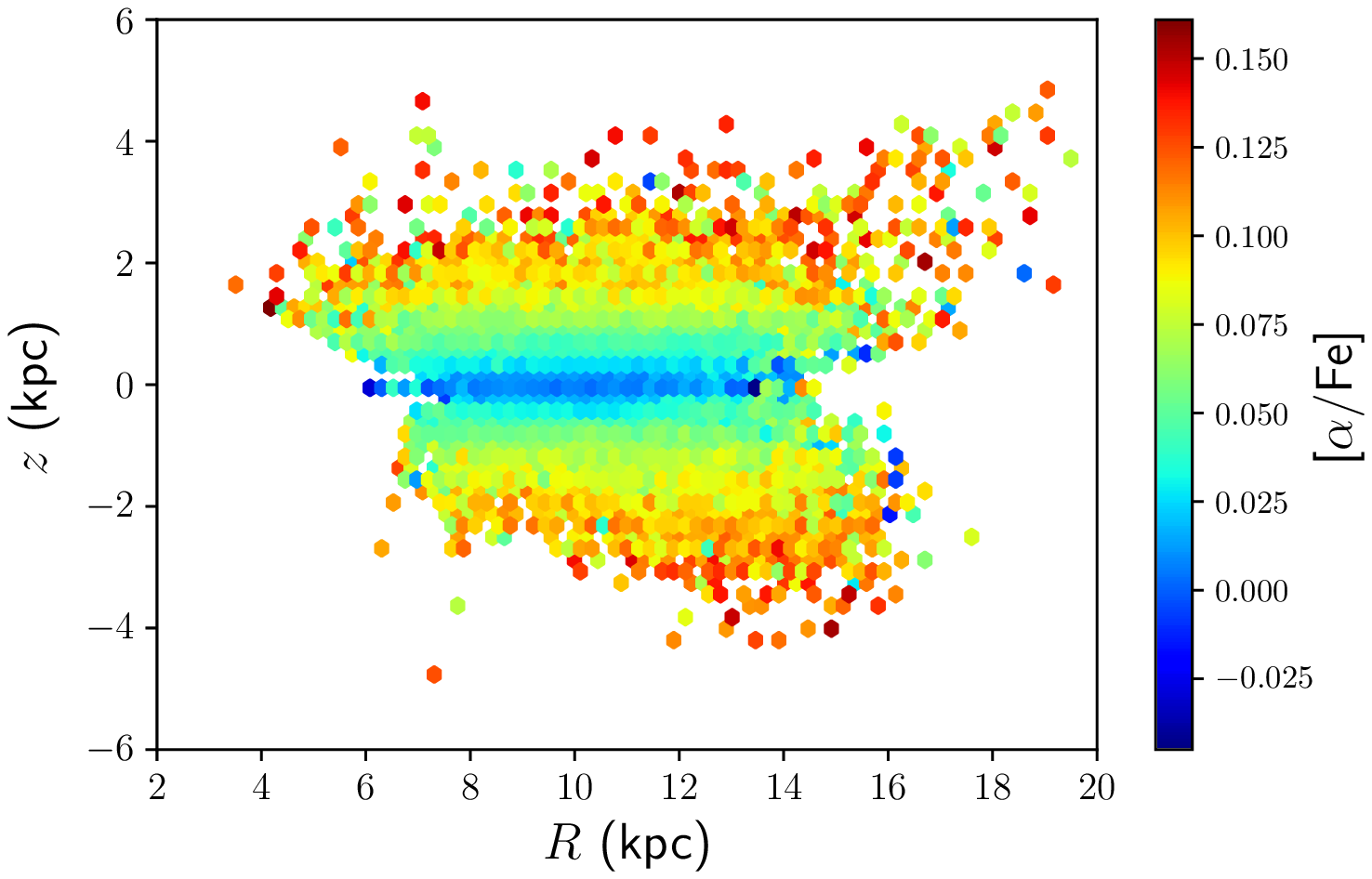}}
    \subfloat[]{\includegraphics[width=1.0\columnwidth]{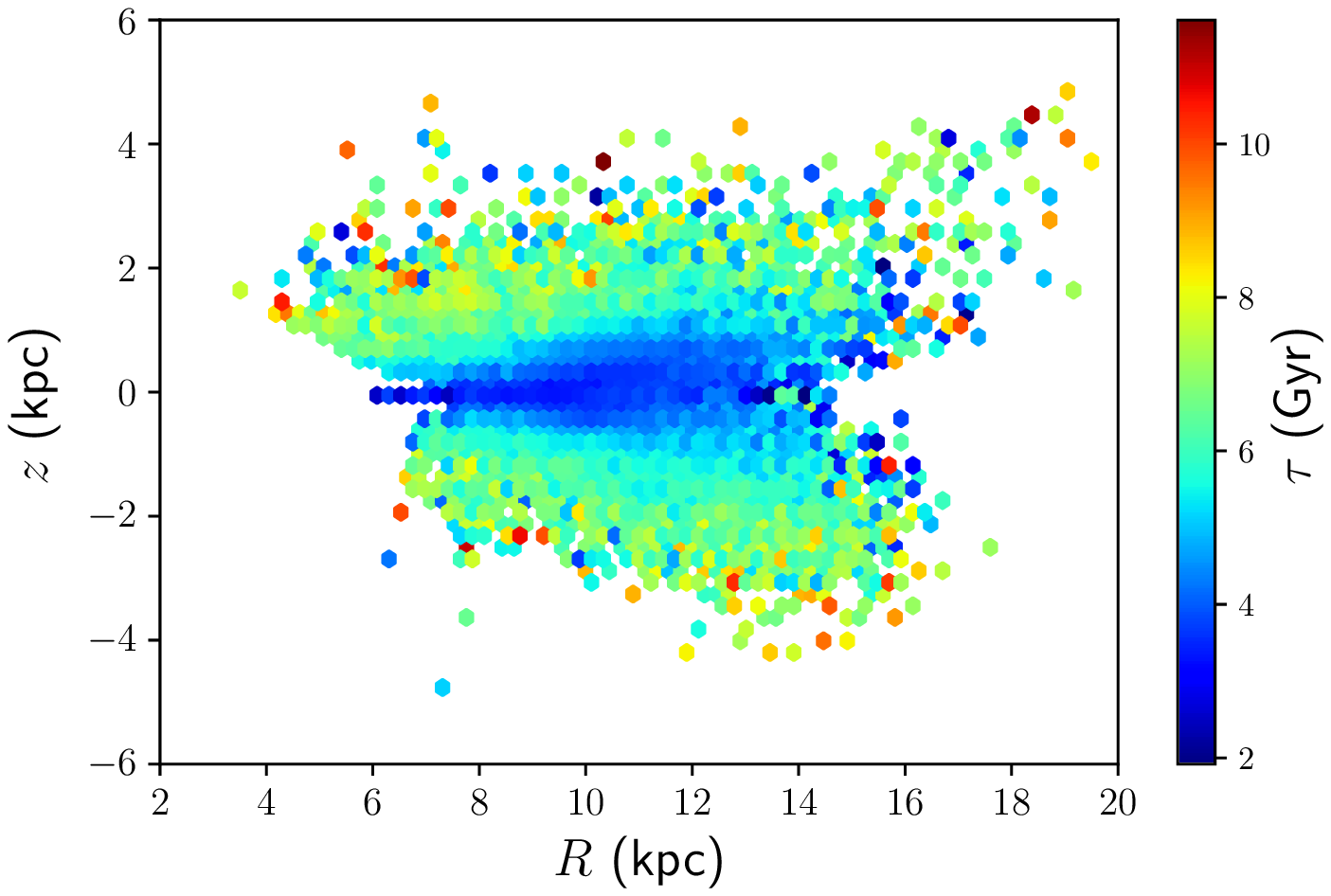}}
    \newline
    \caption{The low-$\alpha$ primary RC sample in the $R$-$z$ plane. The top left panel (a) shows the log-density map, the top right panel (b) shows the $\FeH$ map, the bottom left (c) shows the $\aFe$ map, and the bottom right (d) shows the age map. The sample extends roughly between $3<R<16$ kpc and $-6<z<6$ kpc. The age distribution clearly highlights a young, flaring disc, whose vertical height increases outwards and with age.}\label{fig:data}
\end{figure*}

\section[]{Phase-space, chemical, and age structure}\label{sec:struct}
Here, we explore the structure of the data in phase space, chemistry and age. We adopt a left-handed coordinate system in which positive $v_R$ points away from the Galactic centre and positive $v_{\phi}$ points in the direction of Galactic rotation. To convert from Equatorial coordinates to Galactocentric coordinates, we assume that the Sun is located at $(R_0,z_0) = (8.2,0.014)\kpc$ \citep{mcmillan17}, that the Local Standard of Rest (LSR) has an azimuthal velocity of $232.8\kms$, and that the velocity of the Sun relative to the LSR is $(v_R,v_{\phi},v_z) = (-8.6, 13.9,7.1)\kms$ \citep{mcmillan17}.

\subsection[]{The data}\label{ssec:data}
\cite{huang+20} select stars with spectral signal-to-noise ratios higher than 20 from the 4$^{\mathrm{th}}$ release of the LAMOST Galactic spectroscopic surveys. The selection is based on their positions in the metallicity-dependent effective temperature–surface gravity and color–metallicity diagrams, supervised by high-quality Kepler asteroseismology data. 

The stellar masses and ages of the primary RC stars are determined using the Kernel Principal Component Analysis method, trained with thousands of RCs in the LAMOST-Kepler fields with accurate asteroseismic mass measurements. Typical uncertainties are $15$ and $30\%$, respectively. The purity and completeness of their primary RC sample are generally higher than $80\%$. Using over ten thousand primary RCs with accurate distance measurements from parallaxes provided in the $2^{\mathrm{nd}}$ data release from Gaia, they re-calibrate the $K_{\mathrm{s}}$ absolute magnitudes of primary RCs by considering both the metallicity and age dependencies. With the the new calibration, distances are derived for all the primary RCs, with a typical uncertainty of $5–10\%$. 

We further remove stars from this sample that are likely to be massive, merged stars using the following cut:

\begin{equation}
\aFe > 0.12 \, \mathrm{and} \, \tau < 5.
\end{equation}

\subsection[]{Position, chemistry, and age structure}\label{ssec:pos_chem_age}
Figure \ref{fig:afefehplane} shows the positions of the primary RC sample in the $\aFe$-$\FeH$ plane. The orange line shows a `by-eye' division between the low- and high-$\alpha$ populations. To create our low-$\alpha$ primary RC sample, we select all stars lying below the orange line, which results in a sample of $100,998$ stars.

\begin{figure*}
    \centering
   \includegraphics[width=2.0\columnwidth]{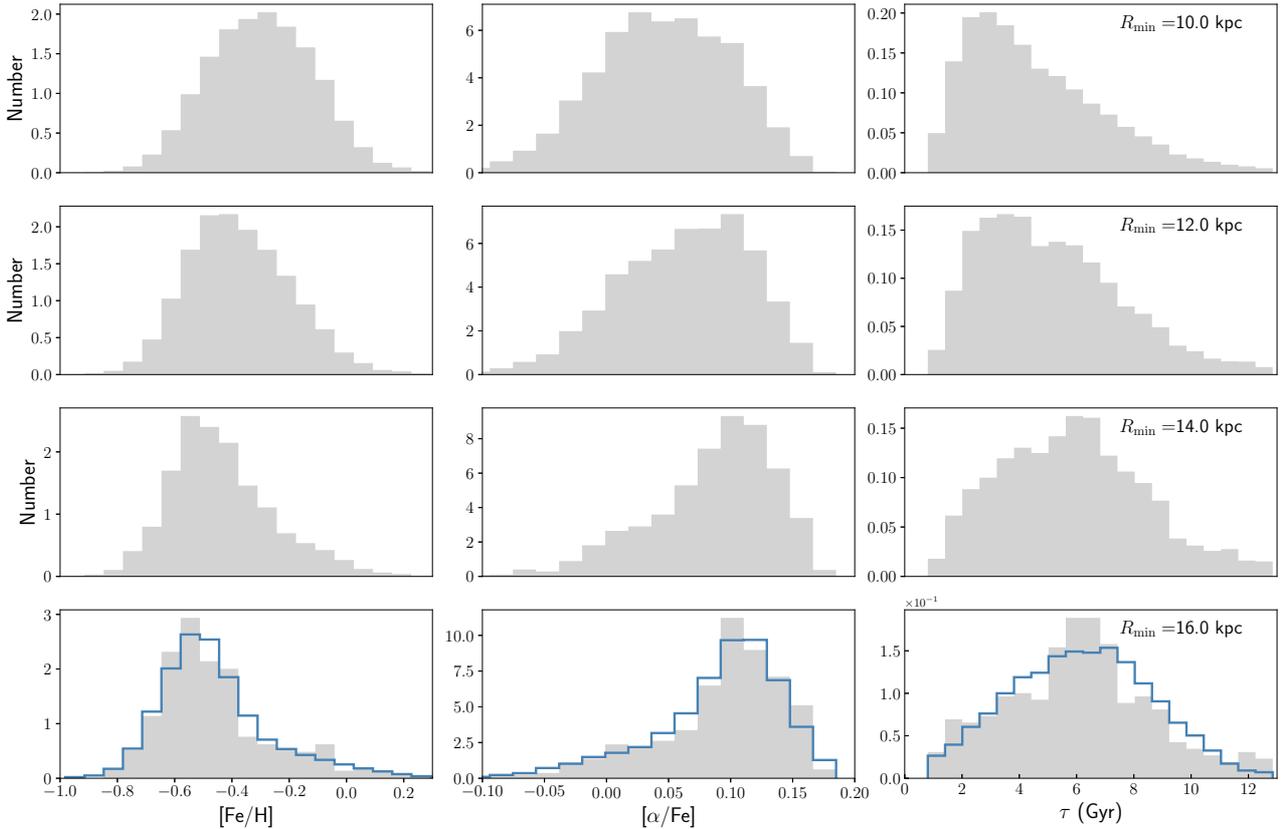}
    \caption{One-dimensional distributions of the low-$\alpha$ primary RC sample in $\FeH$ (left panel), $\aFe$ (middle panel), and $\tau$ (right panel) for stars beyond $10\,\text{kpc}$ (top panel), beyond $12\,\text{kpc}$ (top middle panel), beyond $14\,\text{kpc}$ (bottom middle panel), and beyond $16\,\text{kpc}$ (bottom panel). A metal-poor, $\alpha$-rich, component around $6\,\text{Gyr}$ emerges in the very outer disc. The blue histograms show the two-component three-dimensional GMM fit to the data beyond $16\,\text{kpc}$.}\label{fig:1Dfehafeage}
\end{figure*}

Figure \ref{fig:data} shows the distribution of the low-$\alpha$ primary RC stars in the $R$-$z$ plane. The top left plot shows the density map in this plane. The stars reach as far in as $R=3$ kpc and far out as $R=20$ kpc at higher $z$. The sample extends to $z\sim6$ kpc above and below the mid-Galactic plane. The top right plot shows the $\FeH$ map. The metal richest stars are in the Galactic plane and closer towards the Galactic centre. There is an asymmetry above and below the plane between $12$ and $14\,\text{kpc}$ that likely reflects the lines-of-sight selection function evident in the density map (the selection function of the sample will depend on sky positions, colour, and magnitude, and therefore also age and metallicity). The bottom left panel shows the $\aFe$ map. The stars lowest in $\aFe$ sit in the Galactic plane. There is no gradient with radius. The bottom right plot shows the age map of the stars, clearly highlighting a young, flaring disc, whose vertical height increases with age and Galactocentric distance. Even though $\aFe$ is often considered to be a chemical clock, the flare in age is not seen in $\aFe$, perhaps reflecting different rates of chemical evolution with vertical height.  

Figure \ref{fig:1Dfehafeage} plots one-dimensional distributions of the low-$\alpha$ primary RC stars in $\FeH$, $\aFe$, and age, for stars beyond $10\, \text{kpc}$, $12\,\text{kpc}$, $14\,\text{kpc}$, and $16\,\text{kpc}$. A bump emerges that is metal poor, $\alpha$ rich, and about $6\,\text{Gyr}$ old, seemingly distinct from another component that is broader in $\FeH$, $\aFe$, and $\tau$. We fit a two-component three-dimensional Gaussian Mixture Model (GMM) to the data beyond $16\,\text{kpc}$ finding a narrow component with a mean $\FeH = -0.53$, $\aFe = 0.11$, and $\tau = 7.04\,\text{Gyr}$ and a broad component with a mean $\FeH = -0.27$, $\aFe = 0.04$, and $\tau = 3.84\,\text{Gyr}$.
\begin{table*}
 \centering
  \caption{Names, units, values, and descriptions for parameters of the gravitational potential.\label{tab:gravitationalpotential}}  
	\begin{tabular}{llll}
  	\hline
  	     		 &Parameter/Units   &Value    &Description\\
  	\hline
  	Thin disc 	 &$\Sigma_{0,{\mathrm{thn}}}$/($\mathrm{M}_{\odot} \kpc^{-2}$) &$8.96\times10^8$  &Central surface density\\ 
  	             &$z_{\mathrm{d,thn}}$/$\kpc$                                  &$0.3$             &Scale height\\
  	             &$R_{\mathrm{d,thn}}$/$\kpc$                                  &$2.5$             &Scale radius\\
  	 \hline
  	Thick disc	 &$\Sigma_{0,{\mathrm{thk}}}$/($\mathrm{M}_{\odot} \kpc^{-2}$) &$1.83\times10^8$  &Central surface density\\ 
  	             &$z_{\mathrm{d,thk}}$/$\kpc$                                  &$0.9$             &Scale height\\
  	             &$R_{\mathrm{d,thk}}$/$\kpc$                                  &$3.$              &Scale radius\\
  	 \hline
  	HI gas disc	 &$\Sigma_{0,{\mathrm{HI}}}$/($\mathrm{M}_{\odot} \kpc^{-2}$)  &$5.31\times10^7$ &Central surface density\\ 
  	             &$z_{\mathrm{d,HI}}$/$\kpc$                                 &$0.085$          &Scale height\\
  	             &$R_{\mathrm{d,HI}}$/$\kpc$                                 &$7.0$            &Scale radius\\
  	             &$R_{\mathrm{m,HI}}$/$\kpc$                                 &$4.0$            &Scale radius of hole\\
  	 \hline
  	HII gas disc &$\Sigma_{0,{\mathrm{HII}}}$/($\mathrm{M}_{\odot} \kpc^{-2}$)  &$2.18\times10^9$ &Central surface density\\ 
  	             &$z_{\mathrm{d,HII}}$/$\kpc$                                 &$0.045$          &Scale height\\
  	             &$R_{\mathrm{d,HII}}$/$\kpc$                                 &$1.5$            &Scale radius\\
  	             &$R_{\mathrm{m,HII}}$/$\kpc$                                 &$12.0$           &Scale radius of hole\\
  	 \hline
  	Bulge        &$\rho_{0,{\mathrm{b}}}$/($\mathrm{M}_{\odot} \kpc^{-3}$)      &$9.84\times10^{10}$ &Central surface density\\ 
  	             &$r_{\mathrm{0,b}}$/$\kpc$                                   &$0.075$             &Scale radius\\
  	             &$q_{\mathrm{b}}$                                            &$0.5$               &Axis ratio\\
  	             &$\beta_{\mathrm{b}}$                                        &$0.0$               &Inner slope exponent\\
  	             &$\gamma_{\mathrm{b}}$                                       &$1.8$               &Outer slope exponent\\ 
  	             &$r_{t,\mathrm{b}}$/$\kpc$                                   &$2.1$               &Truncation radius\\  
  	 \hline
  	Dark matter halo &$\rho_{0,{\mathrm{h}}}$/($\mathrm{M}_{\odot} \kpc^{-3}$)  &$8.54\times10^{6}$   &Central surface density\\ 
  	             &$r_{\mathrm{0,h}}$/$\kpc$                          &$19.6$              &Scale radius\\
  	             &$q_{\mathrm{h}}$                                            &$1.0$               &Axis ratio\\
  	             &$\beta_{\mathrm{h}}$                                        &$1.0$               &Inner slope exponent\\
  	             &$\gamma_{\mathrm{h}}$                                       &$3.0$               &Outer slope exponent\\ 
  	             &$r_{t,\mathrm{h}}$/$\kpc$                                   &$\infty$            &Truncation radius\\     

 \hline
  \end{tabular}
\end{table*}

\section[]{Robust vertical velocity dispersion profile}\label{sec:sigmaz}
\begin{figure*}
    \centering
    \subfloat[]{\includegraphics[width=1.0\columnwidth]{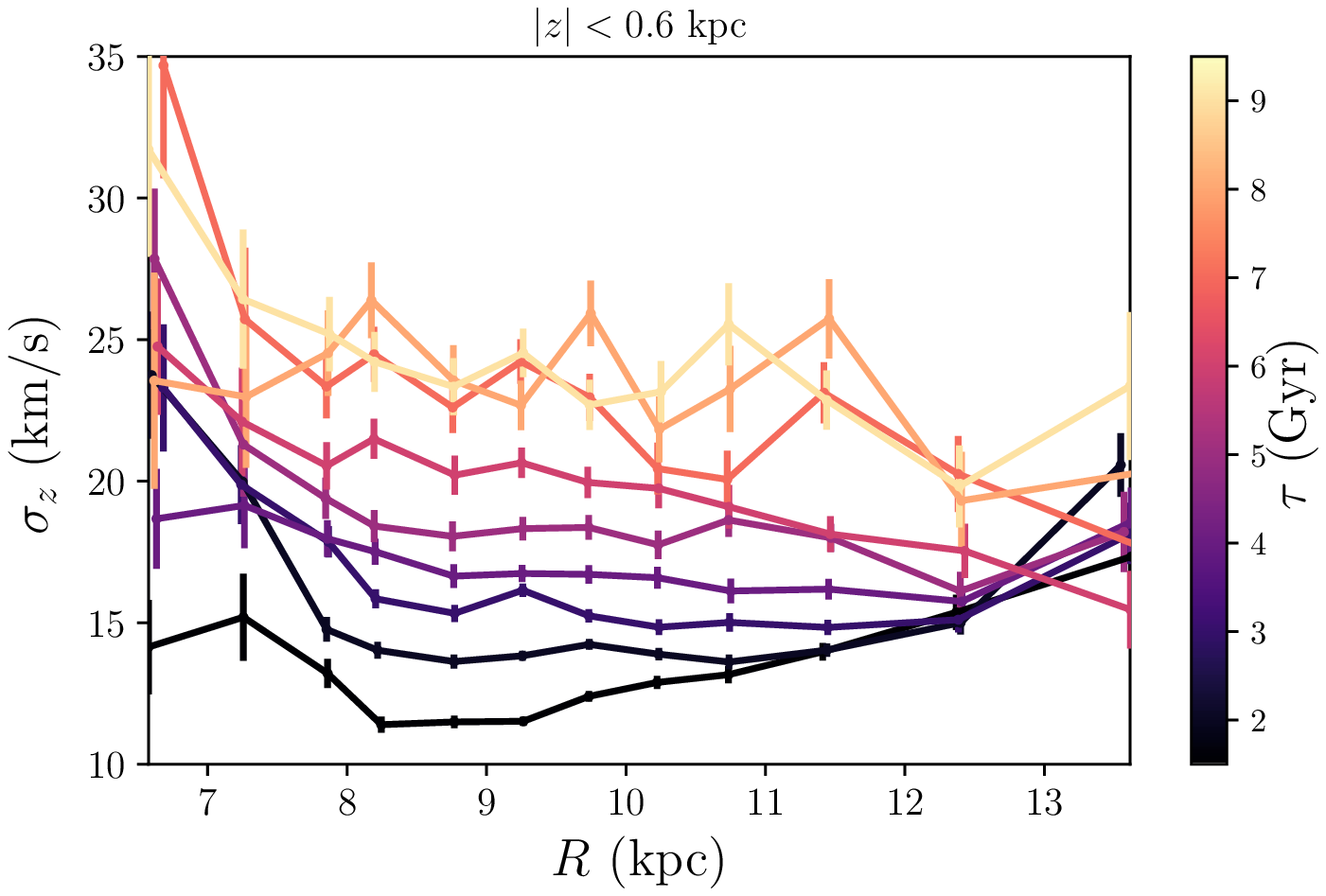}}
    \subfloat[]{\includegraphics[width=1.0\columnwidth]{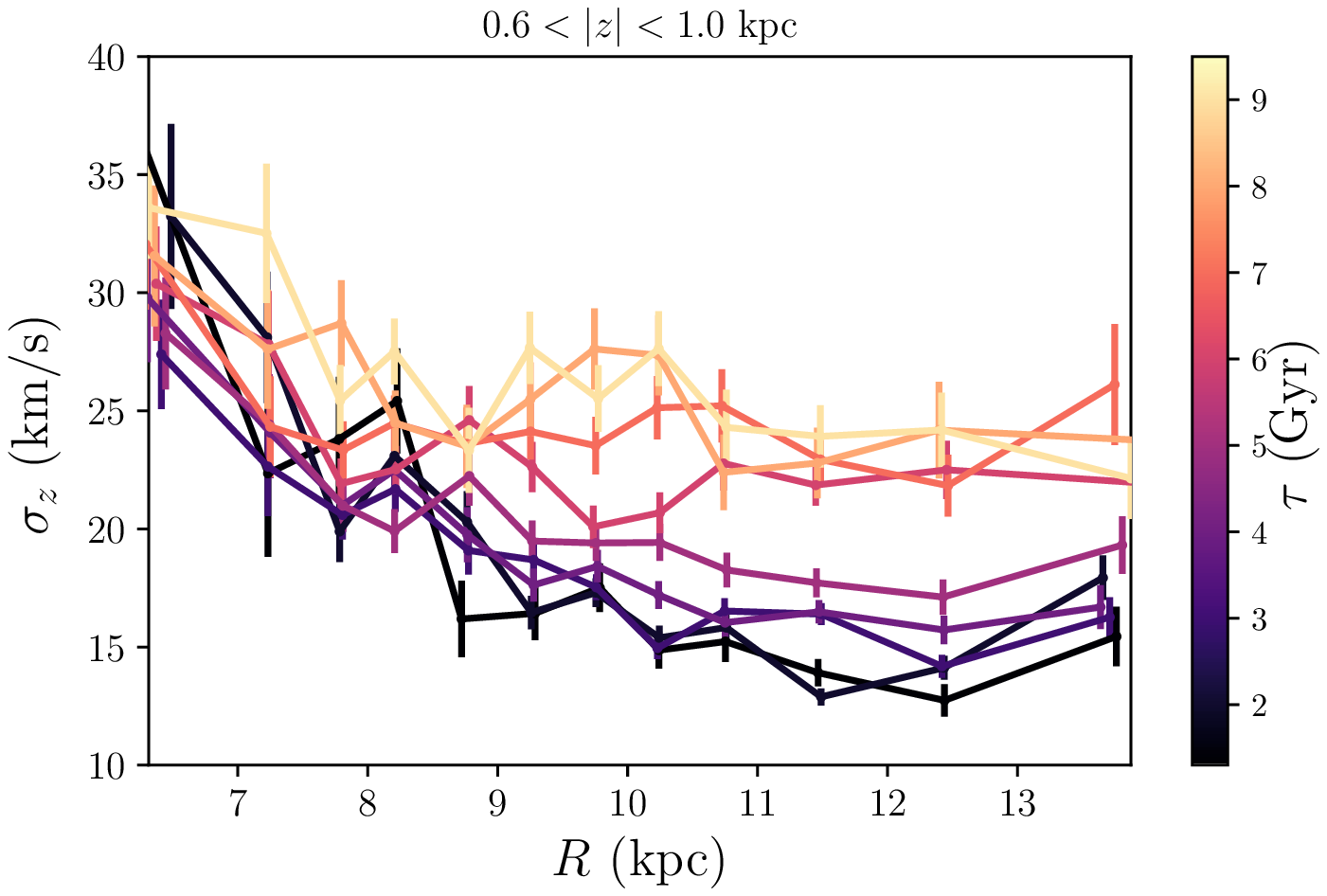}}
    \newline
    \caption{Robust vertical velocity dispersion profiles by age for the low-$\alpha$ primary RC sample in a) the Galactic plane ($|z| < 0.6\,\text{kpc}$) and b) above and below the Galactic plane ($0.6<|z|<1.0\,\text{kpc}$). All profiles decrease within $\sim8\,\text{kpc}$ before plateauing, with the oldest stars having higher vertical velocity dispersions. Beyond $\sim 10\,\text{kpc}$, the vertical velocity dispersion profiles for the youngest stars rise. Note the jump in vertical velocity dispersion at $6\,\textrm{Gyr}$ for stars beyond the Galactic plane.}\label{fig:sigz}
\end{figure*}
We generate the vertical velocity dispersion profile of the low-$\alpha$ primary RC stars, adopting the procedure in \cite{sanders+18}, which allows uncertainties in the velocities to be taken into account. We compute the vertical velocity dispersion of stars in the Galactic plane ($|z| < 0.6\, \text{kpc}$) and just below and above the Galactic plane ($0.6<|z|<1.0\,\text{kpc}$), separated into $12$ radial and $9$ age bins in each case. When computing the velocity dispersion in each bin, we account for the uncertainty in each velocity measurement $\sigma_{vi}$ in the following way. We first perform a $5\sigma$ clip of the raw velocities $v_i$ to remove outliers and halo contaminants. We then seek to maximize the log-likelihood
\begin{equation}
\ln \mathcal{L} = - \frac{1}{2}\sum_i\left(\frac{(v_i-m)^2}{\sigma^2+\sigma_{vi}} + \ln(\sigma^2 + \sigma_{v_i}^2)\right),
\end{equation}

which can be found by solving
\begin{equation}
0 = \sum_i \left( \frac{(v_i - m)^2}{(\sigma^2 + \sigma_{vi}^2)^2} - \frac{1}{\sigma^2 + \sigma_{vi}^2}\right).
\end{equation}

Here, $m$ is the simple mean computed without accounting for the uncertainties. We find the value of $\sigma$ that solves this equation by Brent’s method. We estimate the uncertainty by $\sigma \sqrt{1/2N}$ for $N$ stars. 

In the Galactic plane (Figure \ref{fig:sigz} (a)), the vertical velocity dispersion profiles decrease within $\sim8$ kpc before plateauing, with the oldest stars having higher vertical velocity dispersions. Beyond $\sim10\,\text{kpc}$ the vertical velocity dispersion profiles for the youngest stars rise. Above and below the Galactic plane (Figure \ref{fig:sigz} (b)), the vertical velocity dispersion decreases for all stars until $\sim8\,\text{kpc}$. For stars older than $\sim 6\, \text{Gyr}$ the vertical velocity dispersion profile plateaus between $\sim20$ to $30 \,\text{km/s}$. The vertical velocity dispersion profile of younger stars continues to fall outwards until $\sim 11.5\,\text{kpc}$, where it plateaus. The vertical velocity dispersion profile of stars beyond the Solar Neighbourhood thus appears to `jump' at $\sim 6\, \text{Gyr}$. The mean jump amplitude for $6\text{-Gyr}$ stars beyond $\sim11\, \kpc$ is $\sim 4\,\kms$, greater than the mean error in vertical velocity dispersion of $\sim 1\,\kms$ in the same region. The jump is likely only visible beyond the Galactic plane because the survey lines of sight of the survey result in a higher density of stars at large radii above and below the plane (see Figure \ref{fig:data} (a).
 
\section{The heating effect of an interaction with a satellite}\label{sec:satellite_interaction}
Here we examine the possibility that a satellite underwent a pericentric passage with the Milky Way $\sim6\, \text{Gyr}$ ago, leading to a starburst in the outer disc producing an overabundance of stars at that time, and heating all pre-existing stars (i.e. stars $6\,\text{Gyr}$ old and older) there. We generate a mock Milky Way model of the galaxy that is able to describe the vertical velocity dispersion of the stars younger than $6\,\text{Gyr}$, use the impulse approximation to derive the velocity kicks in the case of a Plummer sphere satellite, then show the change to the vertical velocity dispersion profiles of the outer disc stars after a range of different impacts of varying size, concentration, and position.

\subsection{A mock Milky Way model}\label{ssec:mock_milky_way}
We generate a mock Milky Way with the \texttt{AGAMA} galaxy modelling software \citep{vasiliev+19}. We use the Milky Way gravitational potential fit in \cite{mcmillan17}, generated by thin and thick stellar discs, a HI gas disc, a molecular hydrogen gas disc, and two spheroids representing the bulge and the dark halo. The parameters for their most likely model is given in Table \ref{tab:gravitationalpotential}. 

The densities of the stellar discs are given by
\begin{equation}
	\rho_\mathrm{d}(R,z) = \frac{\Sigma_0}{2z_\mathrm{d}}\exp\left(- \frac{|z|}{z_\mathrm{d}} -\frac{R}{R_\mathrm{d}}  \right),
\end{equation}
with central surface density $\Sigma_0$, scale height $z_\mathrm{d}$, and scale length $R_\mathrm{d}$. The total mass of each stellar disc is $M_d = 2\pi\Sigma_0R_d^2$.

The densities of the gas discs are given by
\begin{equation}
	\rho_\mathrm{d}(R,z) = \frac{\Sigma_0}{4z_\mathrm{d}}\exp\left(- \frac{R_m}{R} -\frac{R}{R_\mathrm{d}}  \right) {\sech}^2(z/2z_d),
\end{equation}
with central surface density $\Sigma_0$, scale height $z_\mathrm{d}$, and scale length $R_\mathrm{d}$. It also has a hole with scale length $R_\mathrm{d}$. The mass of each gas disc is given by $M_d = 2\pi\Sigma_0R_0R_mK_2(2\sqrt{R_m/R_d})$, where $K_2$ is a modified Bessel function.

The densities of the bulge and dark matter halo are given by
\begin{equation}
	\rho(R,z) = \frac{\rho_0}{x^{\gamma}(1+x/r_0)^{\beta-\gamma}}\exp\left[-(x/r_t)^2\right]
\end{equation}
where 
\begin{equation}
	x(R,z) = \sqrt{R^2 + (z/q)^2}. 
\end{equation}
$\rho_0$ sets the density scale, $r_0$ is a scale radius, and the parameter $q$ is the axis ratio of the isodensity surfaces. The exponents $\beta$ and $\gamma$ control the inner and outer slopes of the radial density profile, and $r_\mathrm{t}$ is a truncation radius. 
\begin{table}
 \centering
  \caption{Parameter name and units, values, and descriptions for parameters of the thin-disc distribution function. A dagger indicates the parameter values taken from \protect\cite{piffl+14}. The remaining parameters were fit by eye. \label{tab:df}}  
 \begin{tabular}{lll}
  	\hline
  	Par/Units   &Value    &Description\\
  	\hline 
        $R_d\kpc$                       &$\dagger2.68$     &Scale radius\\
        $\tilde{\sigma}_r/\kms$         &$327.5$           &Central radial velocity dispersion\\
        $\tilde{\sigma}_z/\kms$         &$101.6$           &Central vertical velocity dispersion\\
        $R_{\sigma,r}/\kpc$             &$3.4$             &Radial velocity dispersion scale length\\
        $R_{\sigma,z}/\kpc$             &$5.4$             &Vertical velocity dispersion scale length\\
        $\sigma_{\mathrm{min}}/\kms$    &$1.0$             &Minimum value of velocity dispersion\\
        \hline
  \end{tabular}
\end{table}

For the phase-space distribution of the stars, we assume the action-based quasi-isothermal distribution function (DF) fit to the thin disc in \cite{piffl+14}. The DF for a single population is given by:
\begin{align}
f(\mathbf{J}) &= \frac{\tilde{\Sigma}\Omega}{2\pi^2\kappa^2} \times \frac{\kappa}{\tilde{\sigma}_r^2}\exp \left(-\frac{\kappa J_r}{\tilde{\sigma}_r^2}\right) \times \frac{\nu}{\tilde{\sigma}_z^2} \exp \left(-\frac{\nu J_z}{\tilde{\sigma}_z^2}\right) \times \mathcal{L},\\
\mathcal{L}(J_{\phi}) &\equiv 
  \begin{cases}
   1        & \text{if } J_{\phi} \ge 0, \\
   \exp\left(\frac{2\Sigma J_{\phi}}{\tilde{\sigma}_r^2}\right)        & \text{if } J_{\phi} < 0.
  \end{cases}
  \\
\tilde{\Sigma}(R_c) &\equiv \Sigma_0\exp(-R_c/R_d),\\
\tilde{\sigma}_r^2 &\equiv \tilde{\Sigma}_0\exp(-2R_c/R_{\sigma,r}) + \sigma_{\mathrm{min}^2},\\
\tilde{\sigma}_z^2 &\equiv \Sigma_0\exp(-2R_c/R_{\sigma,z}) + \sigma_{\mathrm{min}^2},
\end{align}
with overall normalization $\tilde{\Sigma}_0$ of the surface density profile. $R_{\mathrm{d}}$ is the disc scale length. The radial velocity dispersion profile is nearly exponential in radius with central value $\sigma_{r,0}$, and radial scale length $R_{\sigma,r}$. The vertical velocity dispersion is also nearly exponential in radius, with central value $\sigma_{z,0}$, and radial scale $R_{\sigma,z}$. $\sigma_{\mathrm{min}}$ is the minimum value of velocity dispersion, added in quadrature to both $\tilde{\sigma}_r$ and $\tilde{\sigma}_z$ to allow sensible behaviour at $J_r=J_z=0$ and large $J_{\phi}$. $R_c$ is the radius of the circular orbit with angular momentum $J_{\phi}$.

\begin{figure}
    \centering
    \subfloat[]{\includegraphics[width=0.85\columnwidth]{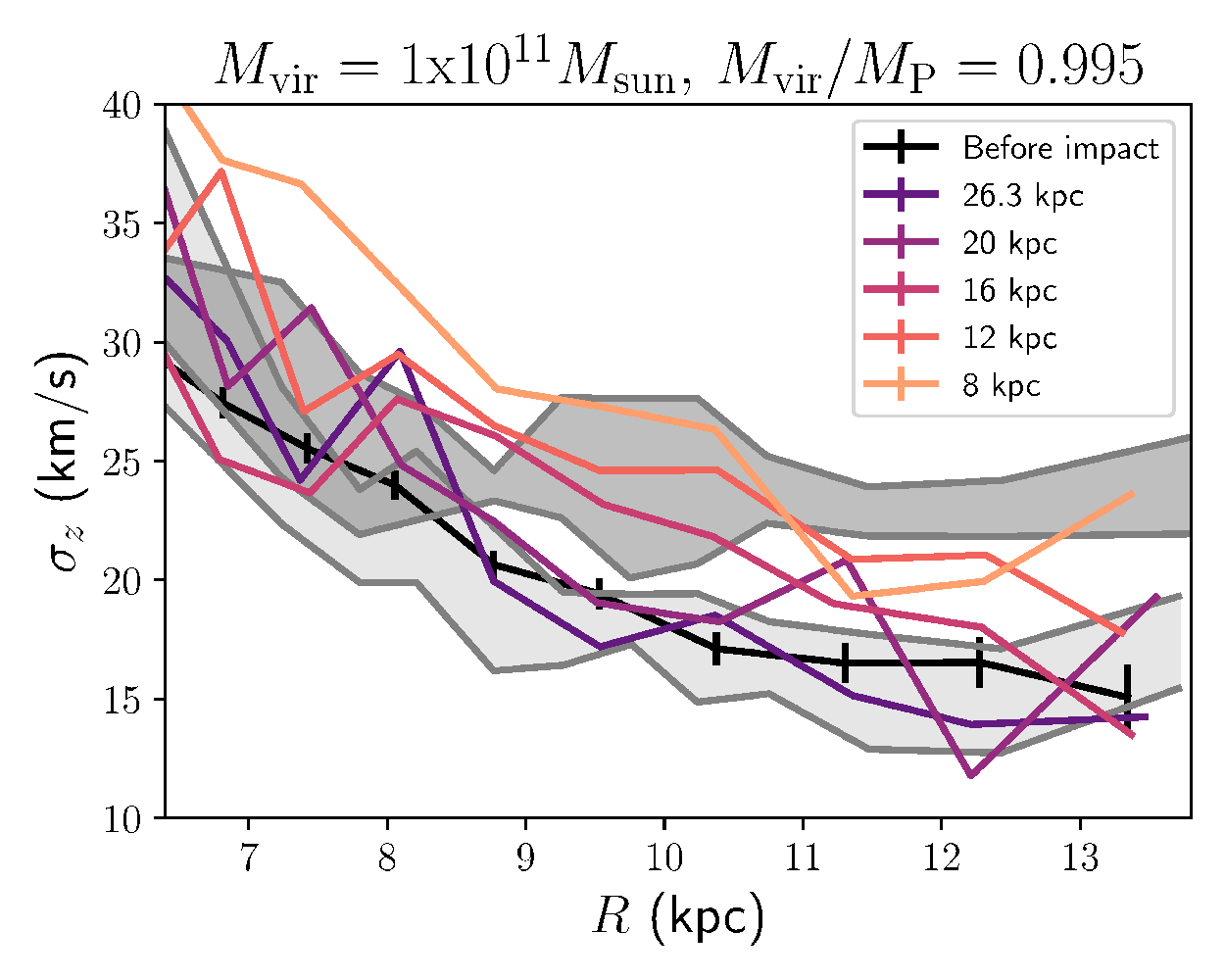}}
    \newline
    \subfloat[]{\includegraphics[width=0.85\columnwidth]{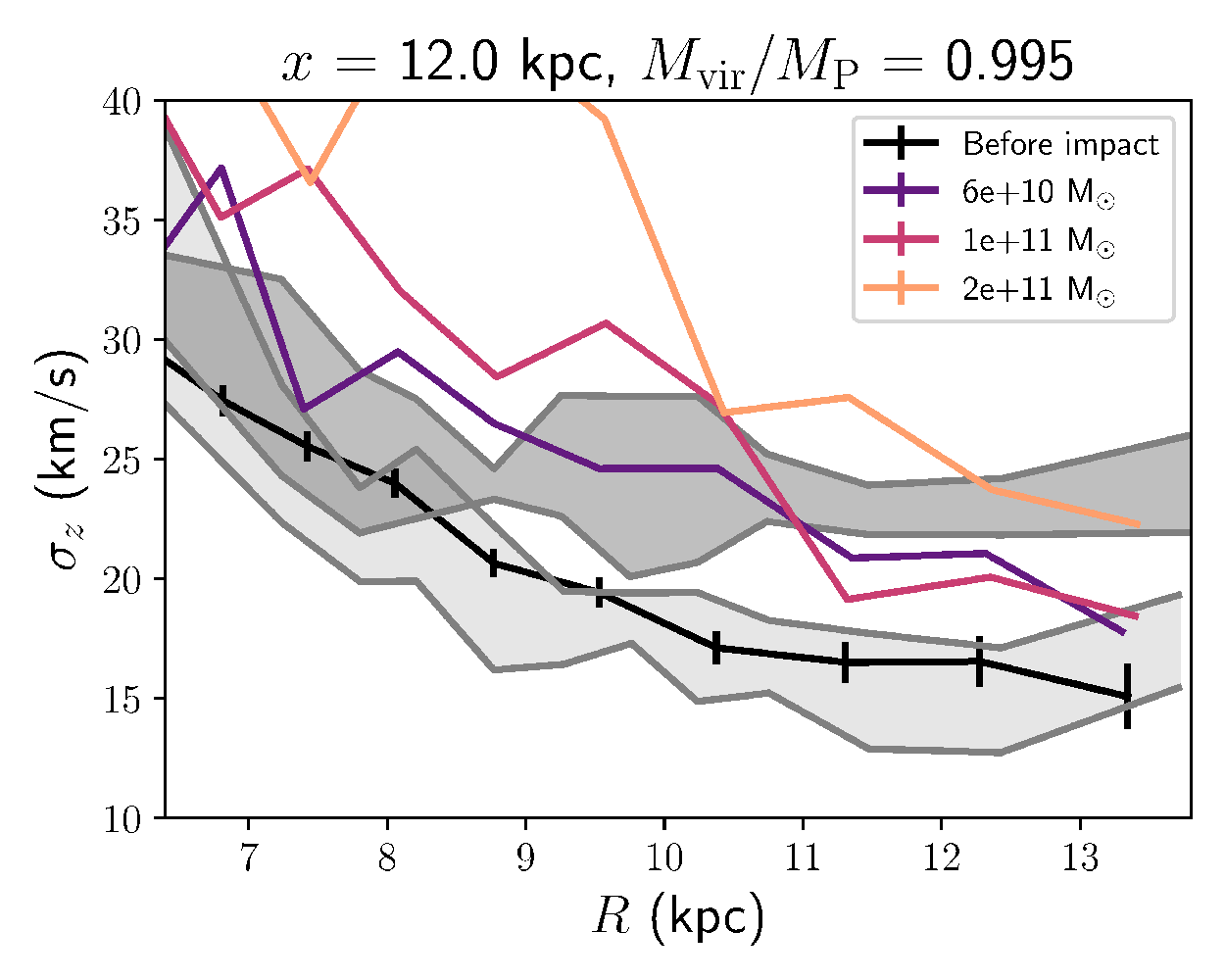}}
    \newline
    \subfloat[]{\includegraphics[width=0.85\columnwidth]{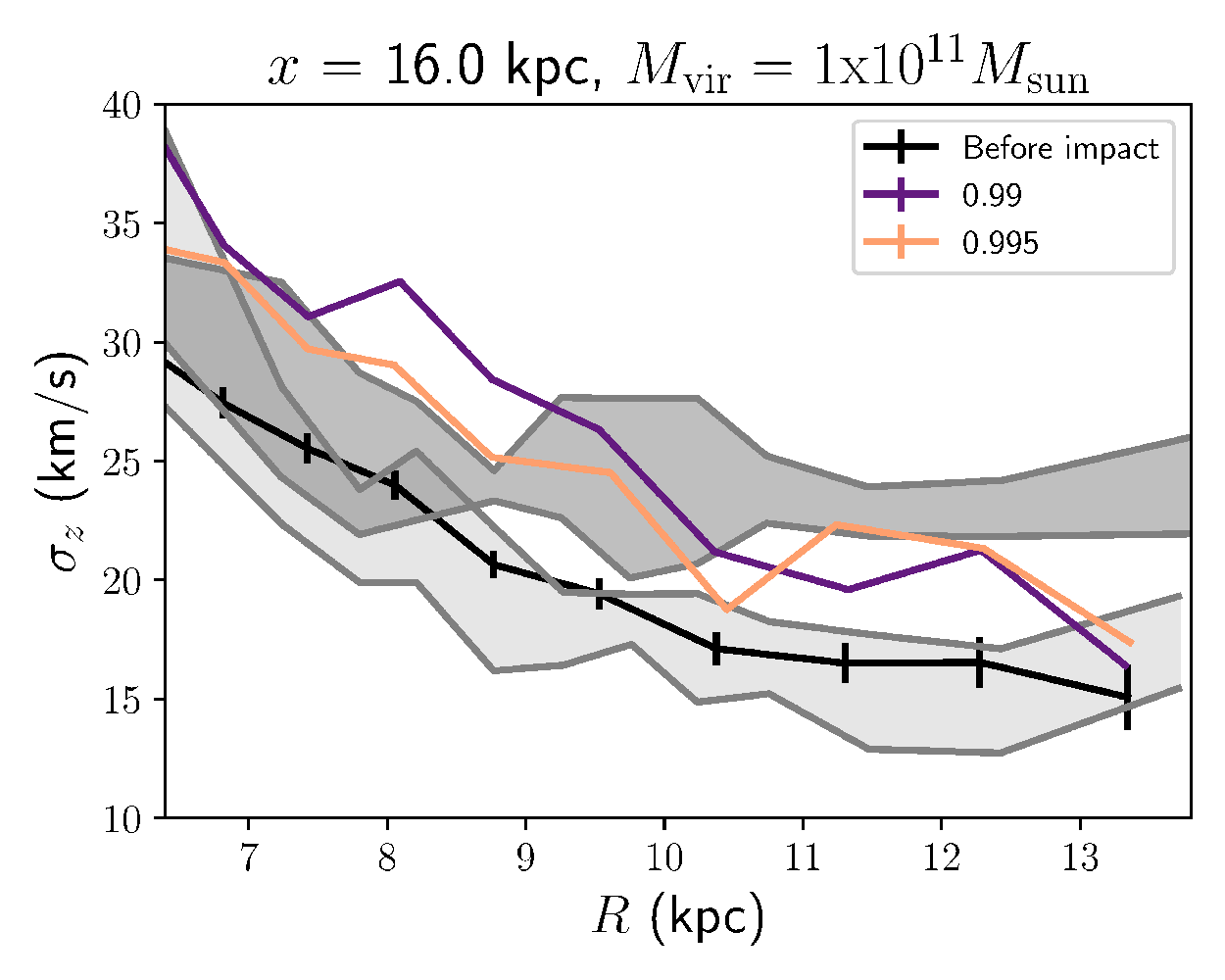}}
    \newline
    \caption{Robust vertical velocity dispersion profiles for stars younger than $6\,\text{Gyr}$ (light grey region), stars older than $6\,\text{Gyr}$, the mock Milky Way model (solid black line), and the mock Milky Way model heated by impacts with (a) an object of virial mass $1 \times 10^{11} M_{\odot}$, virial mass to Plummer mass ratio of $0.995$, and the $x$ component of the impact parameter between $8$ and $26.3\,\text{kpc}$; (b) an object of virial mass beween $6 \times 10^{10} M_{\odot}$ and $2 \times 10^{11} M_{\odot}$, virial mass to Plummer mass ratio of $0.995$, and $x = 12\,\text{kpc}$; and (c) an object of virial mass $1 \times 10^{11} M_{\odot}$, virial mass to Plummer mass ratio of $0.99$ and $0.995$, and $x = 16\,\text{kpc}$. \label{fig:sigz_mock_mw}}
\end{figure}

We generate a sample of $100,000$ stars from the Milky Way model and assign a velocity error of $0.5\,\text{km/s}$ to each coordinate of every star's velocity. We recalculate a robust vertical velocity dispersion profile for stars between an absolute vertical height of $0.6$ and $1.0\,\text{kpc}$, as for the stars in plot (b) of Figure \ref{fig:sigz}, leaving a sample of $12,344$ stars. We modify the parameters of the DF in \cite{piffl+14}, which was fit to data for stars in the Solar Neighbourhood, until we roughly recover the vertical velocity dispersion profile observed for the primary RC stars used in this work for stars younger than $6\,\text{Gyr}$. We take the velocity dispersion profile of stars younger than $6\,\text{Gyr}$ to reflect the vertical velocity dispersion profile of the older stars before the impact with the satellite. The selected DF parameters are given in Table \ref{tab:df}. Figure \ref{fig:sigz_mock_mw} shows the vertical velocity dispersion profile of the mock Milky Way model compared to the vertical velocity dispersion profiles by age for the primary RC stars.

\subsection{Scattering by a Plummer sphere}\label{ssec:vkicks_plummer_sphere}
To determine the velocity kicks imparted to a star in the Milky Way, we consider the scattering impact of a Plummer sphere. We will make the impulse approximation, which is appropriate in situations where external forces on the system are present, but they are small compared to the large, brief internal forces between parts of the system. The gravitational potential of the Plummer sphere, $\Phi$, is given by:
\begin{equation}
\Phi(r) = \frac{GM_0}{\sqrt{r^2 + r_s^2}},
\end{equation}
where $G$ is the gravitational constant, $M_0$ is the mass of the satellite, $r_s$ is the scale radius of the satellite, and $r^2 = (x^2 + y^2 + z^2)$. $(x,y,z)$ is the displacement vector between the star and the satellite.

$(b_x,b_y,b_z)$ is the displacement vector at the point of closest approach between the star and the satellite, and $(w_x,w_y,w_z)$ is the relative velocity at that point. Making this point the zero-point in time, $x_i = b_i + w_i t$, where $i=x,y,z$. Note that $\vec{b}\cdot\vec{w}=0$, $b^2 = \sum_i b_i^2$, and $w^2 = \sum_i w_i^2$.

The acceleration $a_i$ felt by the disc star is given by
\begin{align}
a_i &= \frac{d\Phi}{dx_i}\\
&= \frac{GM_0x_i}{r_s^2 + \sum_i x_i^2}\\
&= \frac{GM_0(b_i + w_i t)}{[r_s^2 + \sum_i(b_i+w_it)^2]^{3/2}}\\
&= \frac{GM_0(b_i + w_i t)}{[r_s^2 + \sum_i(b_i^2 + w_i^2t^2)]^{3/2}} \\
&= \frac{GM_0(b_i + w_i t)}{(r_s^2 + b^2 + w^2t^2)^{3/2}}
\end{align}

Defining $\chi^2 \equiv r_s^2 + b^2$, the resulting velocity kick is then given by
\begin{align}\label{eqn:impulse}
\begin{split}
    \Delta v_i &= \int_{-\infty}^{\infty} a_i dt \\
    &= \int_{-\infty}^{\infty} \frac{GM_0(b_i + w_i t)}{(\chi^2 + w^2t^2)^{3/2}} dt \\
    &= GM_0\left(\frac{b_i}{w^3} \int_{-\infty}^{\infty} \frac{1}{(\frac{\chi^2}{w^2} + t^2)^{3/2}} dt + \frac{w_i}{w^3} \int_{-\infty}^{\infty} \frac{t}{(\frac{\chi^2}{w^2} + t^2)^{3/2}} dt \right) \\
    &= GM_0\left[\frac{b_i}{w^3}\left(\frac{t}{\frac{\chi^2}{w^2}(\frac{\chi^2}{w^2} + t^2)^{1/2}}\right) - \frac{w_i}{w^3}\left(\frac{\chi^2}{w_i^2} + t^2\right)^{-1/2}\right]_{-\infty}^{\infty}\\
    &= \frac{2GM_0b_i}{w\chi^2}.
\end{split}
\end{align}
The result is a simple expression for the velocity kick.

\subsection{Simulated vertical velocity dispersion profiles following impact}\label{ssec:simulated_sigz}
To simulate the potential impact of a satellite on the outer thin disc of the Milky Way, we explore the possibility that the impact $6\,\text{Gyr}$ ago was with a Plummer-sphere-like object. \cite{vasiliev+20} perform a number of N-body simulations of a disrupting Sgr galaxy as it orbits the Milky Way over the past $2.5\,\textrm{Gyr}$, tailored to reproduce the observed properties of the remnant as measured by the second data release from Gaia \citep{gaia+18} and line-of-sight velocities from other surveys. At the last pericentric passage, they find a relative velocity between the satellite and the Milky Way of $w_{\text{Sag-MW}} = (-45.82, -93.79, 255.52)$ km/s, and impact parameter $b_{\text{Sag-MW}} = (26.27,-0.821,-1.397)\,\text{kpc}$. Here, we assume the same relative velocity and impact parameter, but additionally consider $x = 8,\,12,\,16,\,\text{and}\,20\,\text{kpc}$. We explore virial masses similar to that explored for the progenitor Sgr in \cite{laporte+18}, in their exploration of the influence of Sgr and the Large Magellanic Cloud (LMC) on the stellar disc of the Milky Way. Specifically, we explore virial masses of $6\times10^{10},\,1\times10^{11},\,\,\text{and}\,2\times10^{11} M_{\odot}$. We also explore different concentrations, assuming the virial mass is $0.99,\,\text{and}\,0.995$ of the Plummer mass. This corresponds to different scale radii of the Plummer sphere. We ran $5\times3\times2=30$ simulations in total.

We use Equation \eqref{eqn:impulse} to derive the velocity kicks given the impact parameters for a random sample of $10,000$ stars from the $100,000$ in our mock sample. We then relax the $10,000$ stars for approximately two dynamical timescales (evaluated for the outer disc), about $4\,\text{Gyr}$. We recalculate a robust vertical velocity dispersion profile and re-select stars between an absolute vertical height of $0.6$ and $1.0\,\text{kpc}$. 

Figure \ref{fig:sigz_mock_mw} shows a selection of these results. (a) shows the effect of the impact parameter (in terms of the $x$ coordinate) for a virial mass $1\times10^{11}M_{\odot}$ and a virial mass to Plummer mass ratio of $0.995$. The closer the object to the centre of the Milky Way, the bigger the effect on the vertical velocity dispersion, especially within the solar radius. Increasing the mass of the object (see (b)) also increases the effect, in particular within the solar radius. Increasing the concentration (see (c)) appears to steepen the effect with radius. The simulations that most closely recovered the vertical velocity dispersion of primary RC stars beyond the Galactic Plane older than $6 \, \text{Gyr}$ were the models with a virial mass of $6\times10^{10} M_{\odot}$, $x=12\,\text{kpc}$, and a virial mass to Plummer mass ratio of $0.995$ (shown in Figure \ref{fig:sigz_mock_mw} (b)) and a virial mass of $1\times10^{11} M_{\odot}$, $x=16\,\text{kpc}$, and a virial mass to Plummer mass ratio of $0.995$ (shown in Figure \ref{fig:sigz_mock_mw} (a)).

\section[]{Discussion}\label{sec:discussion}
Here we discuss other interpretations of the chrono-chemodynamical structure of the outer disc, the evidence supporting Sgr as the massive satellite in the case such a scenario is the correct one, and uncertainties in the analysis.

\subsection[]{Other interpretations of the chrono-chemodynamical structure of the outer disc}
There are other possible interpretations of the chrono-chemodynamical structure in the outer disc of the Milky Way. An abrupt jump in the vertical velocity dispersion profile at $6\,\text{Gyr}$ is unlikely the result of background heating by giant molecular clouds (GMCs) or dark matter subhaloes, as these occur continuously. Radial migration from the inner disc is also an unlikely explanation. It is unclear whether the velocity dispersion of the radially-migrated stars should be higher \citep[e.g.][]{veraciro+14} and why the effect would only be discernible for stars older than $6\,\text{Gyr}$. The radially-migrated stars should be metal richer too.

The scenario proposed in \cite{renaud+21b}, where the outer gas disc is fed by a separate gas filament to that feeding the inner gas disc is a possibility. The outer gas disc is initially misaligned with respect to the inner gas disc. Stars form in the outer disc after a significant minor merger event, and then the discs align resulting in a hot outer disc. In this scenario however, the stars born in the outer disc at the time of the significant minor merger event should be the kinematically hot stars. In the RC sample, all stars beyond the Galactic plane, older than $6\,\text{Gyr}$ are hot, even though the overdensity in $\FeH$, $\aFe$, and $\tau$ appears to be centred at around $7\,\text{Gyr}$.

The scenario in which a massive satellite produced the jump in the vertical velocity dispersion profiles at $6\,\text{Gyr}$ is supported by simulations ran by \cite{martig+14b}. They studied the relation between stellar ages and vertical velocity dispersion (age-velocity dispersion relation, AVR) in a sample of seven simulated disc galaxies. They found that the shape of the AVR for stars younger than $9\,\text{Gyr}$ depended strongly on the merger history at low redshift, with even $1:10$-$1:15$ mergers being able to create jumps in the AVR ( these jumps may not be visible if age errors are $\gtrsim 30\%$). The additional finding of an overdensity in the $\FeH$, $\aFe$, $\tau$ hyperplane, suggesting a starburst induced by the pericentric passage of the massive satellite, further supports this scenario.

\begin{figure*}
    \centering
    \includegraphics[width=2.0\columnwidth]{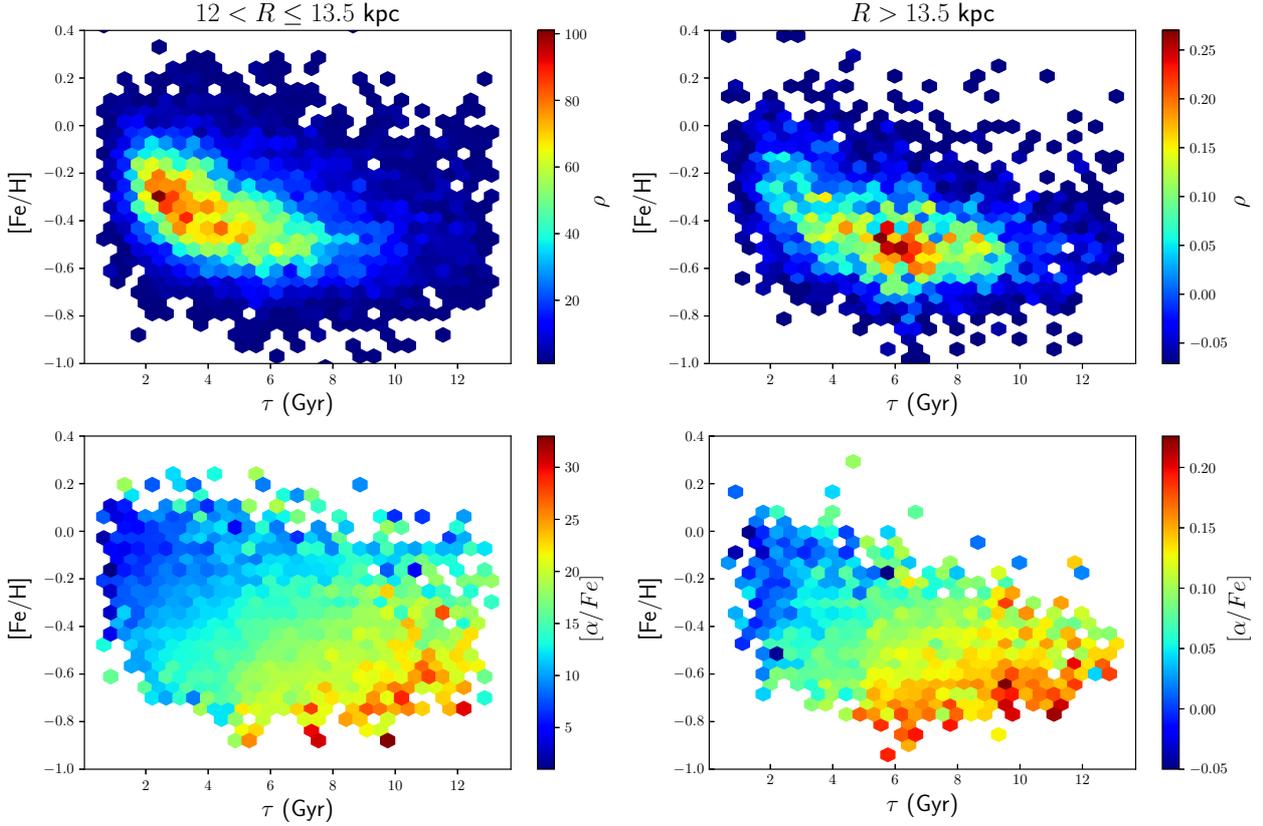}
    \caption{Hexbin plots in the $\FeH$-$\tau$ plane for stars between $12$ and $13.5\,\textrm{Gyr}$ (left panel) and beyond $13.5\,\text{kpc}$ (right panel) showing density maps (top) and $\aFe$ maps (bottom).  Note the overdensity at $\sim 7\,\text{Gyr}$ and $\FeH \sim -0.5$ in the density map for stars beyond $13.5\,\textrm{Gyr}$, and the faint streak towards lower $\FeH$ and higher $\aFe$ just under the overdensity.\label{fig:age_feh_afe_outer_disc}}
\end{figure*}

Mergers and pericentric passages of satellites can bring metal-poorer gas that dilutes the pre-existing gas \citep{grand+20, renaud+21a}. New stars that form would be metal poorer than expected. Pre-existing metal-poorer stars born in the shallower satellite potential could also be deposited in the host. Both scenarios should produce a dip towards lower metallicities at the same age in the $\FeH$-$\tau$ plane \citep{renaud+21a,ciuca+23}. Figure \ref{fig:age_feh_afe_outer_disc} zooms into the $\FeH$-$\tau$ plane for stars just within and beyond $13.5 \textrm{kpc}$. The density map beyond $13.5 \textrm{kpc}$ shows a clear overdensity at ages just below $7\,\text{Gyr}$, and metallicities around $0.5$, corresponding to the second bump seen in Figure \ref{fig:1Dfehafeage}. There is very tentative evidence for a streak towards lower metallicities around the same age. The bottom panels show the $\aFe$ maps in this plane. The faint streak beyond $13.5 \textrm{kpc}$ is also higher in $\aFe$.

\subsection[]{The possibility of Sgr as the massive satellite perturber}
%
We find that an object of mass $\sim 10^{11}$ is required to produce the observed jump in vertical velocity dispersion at $6\,\text{Gyr}$ for primary RC stars above and below the Galactic plane, assuming the object had the same relative velocity, $y$, and $z$ coordinates as Sgr at its last pericentric passage \citep{vasiliev+20}, but $x = 12\,\text{to}\,16\,\text{kpc}$ .

Many attempts exist in the literature at estimating the progenitor mass of Sgr. \cite{niederste_ostholt+10} reassemble the stellar debris using Sloan Digital Sky Survey (SDSS) and Two Micron All-Sky Survey data to estimate a mass for Sgr's dark matter halo, prior to tidal disruption, of $\sim10^{10}M_{\odot}$. \cite{gibbons+17} use SDSS stars in the leading and trailing streams of Sgr to demonstrate the existence of two sub-populations with distinct chemistry and kinematics. They show that that the dark halo of Sgr must have been $\gtrsim 6 \times 10^{10} M_{\odot}$ to reproduce the observed velocity dispersions. \cite{mucciarelli+17} attempt to reproduce the observed chemical patterns of Sgr and conclude that the total progenitor mass is $6 \times 10^{10}M_{\odot}$. The estimate we derive here is very similar to these previous estimates in the literature.

For a system the size of our Milky Way ($\sim 10^{12} M_{\odot}$, e.g. \cite{craig+22}), we only expect one or two $10^{8}$-mass satellites, one $10^{9}$-mass satellite and not really any $10^{10}$-mass satellites \citep{santos+22}. The fact there is a Large Magellanic Cloud (LMC), which may have a mass up to $\sim 10^{11} M_{\odot}$ as well as Sgr, also with a mass up to $\sim 10^{11} M_{\odot}$ is already unusual. It is very unlikely yet another massive satellite exists in the halo of the Milky Way. The LMC is around $50\,\text{kpc}$ \citep{pietr+13} from the centre of the Milky Way, which is too far away to have the observed effect on the vertical velocity dispersion. This only leaves Sgr as the possible candidate.


The pericentric radius inferred in this work $6\,\text{Gyr}$ ago is between $12$ and $16\,\text{kpc}$. This is closer than the $26.3\,\text{kpc}$ found for the last pericentric passage of Sgr in \cite{vasiliev+20}. This is unexpected as one expects large satellites to sink towards the centre of the host galaxy due to dynamical friction. However, recent work analysing the orbital dynamics and histories of satellite galaxies around Milky Way (MW)-mass host galaxies in the FIRE-2 cosmological simulations, found that for $67\%$ of the satellites, the most recent pericentre was not their minimum pericentre \citep{santistevan+22}. The minimum typically was $40\%$ smaller and occurred $\sim 6\,\text{Gyr}$ earlier, possibly coincidentally, very similar to what we find here. They attribute the increasing pericentres to a combination of a time-dependent Milky-Way-mass halo potential and dynamical perturbations in the outer halo.

\section[]{Conclusions and further work}\label{sec:conc}
A primary RC sample drawn from the $4^{\text{th}}$ data release of LAMOST was used to examine the chrono-chemodynamical structure of the Milky Way, with a focus on the outer disc. The vertical velocity dispersion profiles by age for stars beyond the Galactic plane show a jump at $6\,\text{Gyr}$. The chemistry and age histograms reveal a bump at $\FeH = -0.5$, $\aFe=0.1$, and an age of $7.2\,\text{Gyr}$ and stars beyond $13.5\,\text{kpc}$ show a dip in metallicity just below the bump. These data suggest that a massive satellite had a close encounter with the Milky Way $\sim6\,\text{Gyr}$ ago, heating all pre-existing stars, causing a starburst, and bringing in metal-poorer gas that subsequently formed stars. The impulse approximation was used to weigh and determine the orbit of the massive satellite, finding a mass $\sim 10^{11}M_{\odot}$ and a pericentric radius between $12$ and $16\,\text{kpc}$. The massive satellite is likely to be Sgr.

The work shows the potential for the second-order moments of velocity to encode more ancient interactions with massive satellites (rather than the first-order velocity moments that encode more recent interactions, e.g. \cite{laporte+19}).

The second paper in the series will address a key uncertainty in the analysis presented here. The starburst in the outer disc was found to be centred at $\sim 7\,\text{Gyr}$ and $\FeH\sim-0.5$. The primary RC clump sample has a strong sample selection function due to the selection in colour and metallicity. This will result in a bias in the distance-metallicity-age distribution of the stars that needs to be accounted for to correctly position the starburst in the age-metallicity plane. The second paper presents an extended distribution function model for the outer disc that takes the selection function into account.

The second uncertainty in the analysis presented here is in the application of the impulse approximation to estimate the orbital parameters of the satellite. The impulse approximation assumes that the dynamical heating effect is dominated by the velocity kicks imparted on the disc stars by the satellite at the point of closest approach. In future work, we will run N-body simulations to follow this impact and explore a wider parameter space.

\section*{Acknowledgements}
PD is supported by a UKRI Future Leaders Fellowship (grant reference MR/S032223/1). YH is supported by National Key R\&D Program of China No. 2019YFA0405500 and National Natural Science Foundation of China grants 11903027 and 11833006.

This work has made use of data from the European Space Agency (ESA) mission
{\it Gaia} (\url{https://www.cosmos.esa.int/gaia}), processed by the {\it Gaia} Data Processing and Analysis Consortium (DPAC, \url{https://www.cosmos.esa.int/web/gaia/dpac/consortium}). Funding for the DPAC has been provided by national institutions, in particular the institutions participating in the {\it Gaia} Multilateral Agreement.

The Guoshoujing Telescope (the Large Sky Area Multi-Object Fiber Spectroscopic Telescope, LAMOST) is a National Major Scientific Project built by the Chinese Academy of Sciences. Funding for the project has been provided by the National Development and Reform Commission. LAMOST is operated and managed by the National Astronomical Observatories, Chinese Academy of Sciences. The LAMOST FELLOWSHIP is supported by Special fund for Advanced Users, budgeted and administrated by Center for Astronomical Mega-Science, Chinese Academy of Sciences (CAMS). 

\bibliographystyle{mnras}
\bibliography{payeldas}

\begin{thebibliography}{}
\makeatletter
\relax
\def\mn@urlcharsother{\let\do\@makeother \do\$\do\&\do\#\do\^\do\_\do\%\do\~}
\def\mn@doi{\begingroup\mn@urlcharsother \@ifnextchar [ {\mn@doi@}
  {\mn@doi@[]}}
\def\mn@doi@[#1]#2{\def\@tempa{#1}\ifx\@tempa\@empty \href
  {http://dx.doi.org/#2} {doi:#2}\else \href {http://dx.doi.org/#2} {#1}\fi
  \endgroup}
\def\mn@eprint#1#2{\mn@eprint@#1:#2::\@nil}
\def\mn@eprint@arXiv#1{\href {http://arxiv.org/abs/#1} {{\tt arXiv:#1}}}
\def\mn@eprint@dblp#1{\href {http://dblp.uni-trier.de/rec/bibtex/#1.xml}
  {dblp:#1}}
\def\mn@eprint@#1:#2:#3:#4\@nil{\def\@tempa {#1}\def\@tempb {#2}\def\@tempc
  {#3}\ifx \@tempc \@empty \let \@tempc \@tempb \let \@tempb \@tempa \fi \ifx
  \@tempb \@empty \def\@tempb {arXiv}\fi \@ifundefined
  {mn@eprint@\@tempb}{\@tempb:\@tempc}{\expandafter \expandafter \csname
  mn@eprint@\@tempb\endcsname \expandafter{\@tempc}}}

\bibitem[\protect\citeauthoryear{{Am{\^o}res}, {Robin}  \&
  {Reyl{\'e}}}{{Am{\^o}res} et~al.}{2017}]{amores+17}
{Am{\^o}res} E.~B.,  {Robin} A.~C.,   {Reyl{\'e}} C.,  2017, \mn@doi [\aap]
  {10.1051/0004-6361/201628461}, \href
  {https://ui.adsabs.harvard.edu/abs/2017A&A...602A..67A} {602, A67}

\bibitem[\protect\citeauthoryear{{Arenou} et~al.,}{{Arenou}
  et~al.}{2018}]{arenou+18}
{Arenou} F.,  et~al., 2018, \mn@doi [\aap] {10.1051/0004-6361/201833234}, \href
  {https://ui.adsabs.harvard.edu/abs/2018A&A...616A..17A} {616, A17}

\bibitem[\protect\citeauthoryear{{Bovy}, {Rix}, {Schlafly}, {Nidever},
  {Holtzman}, {Shetrone}  \& {Beers}}{{Bovy} et~al.}{2016}]{bovy+16}
{Bovy} J.,  {Rix} H.-W.,  {Schlafly} E.~F.,  {Nidever} D.~L.,  {Holtzman}
  J.~A.,  {Shetrone} M.,   {Beers} T.~C.,  2016, \mn@doi [\apj]
  {10.3847/0004-637X/823/1/30}, \href
  {http://adsabs.harvard.edu/abs/2016ApJ...823...30B} {823, 30}

\bibitem[\protect\citeauthoryear{{Chiang}, {Ostriker}  \& {Schive}}{{Chiang}
  et~al.}{2023}]{chiang+23}
{Chiang} B.~T.,  {Ostriker} J.~P.,   {Schive} H.-Y.,  2023, \mn@doi [\mnras]
  {10.1093/mnras/stac3358}, \href
  {https://ui.adsabs.harvard.edu/abs/2023MNRAS.518.4045C} {518, 4045}

\bibitem[\protect\citeauthoryear{{Church}, {Mocz}  \& {Ostriker}}{{Church}
  et~al.}{2019}]{church+19}
{Church} B.~V.,  {Mocz} P.,   {Ostriker} J.~P.,  2019, \mn@doi [\mnras]
  {10.1093/mnras/stz534}, \href
  {https://ui.adsabs.harvard.edu/abs/2019MNRAS.485.2861C} {485, 2861}

\bibitem[\protect\citeauthoryear{{Ciuc{\u{a}}}, {Kawata}, {Miglio}, {Davies}
  \& {Grand}}{{Ciuc{\u{a}}} et~al.}{2021}]{ciuca+21}
{Ciuc{\u{a}}} I.,  {Kawata} D.,  {Miglio} A.,  {Davies} G.~R.,   {Grand} R.
  J.~J.,  2021, \mn@doi [\mnras] {10.1093/mnras/stab639}, \href
  {https://ui.adsabs.harvard.edu/abs/2021MNRAS.503.2814C} {503, 2814}

\bibitem[\protect\citeauthoryear{{Ciuc{\u{a}}} et~al.,}{{Ciuc{\u{a}}}
  et~al.}{2022}]{ciuca+23}
{Ciuc{\u{a}}} I.,  et~al., 2022, arXiv e-prints, \href
  {https://ui.adsabs.harvard.edu/abs/2022arXiv221101006C} {p. arXiv:2211.01006}

\bibitem[\protect\citeauthoryear{{Craig}, {Chakrabarti}, {Baum}  \&
  {Lewis}}{{Craig} et~al.}{2022}]{craig+22}
{Craig} P.~A.,  {Chakrabarti} S.,  {Baum} S.,   {Lewis} B.~T.,  2022, \mn@doi
  [\mnras] {10.1093/mnras/stac2308}, \href
  {https://ui.adsabs.harvard.edu/abs/2022MNRAS.517.1737C} {517, 1737}

\bibitem[\protect\citeauthoryear{{Das} \& {Sanders}}{{Das} \&
  {Sanders}}{2019}]{das+19}
{Das} P.,  {Sanders} J.~L.,  2019, \mn@doi [\mnras] {10.1093/mnras/sty2776},
  \href {https://ui.adsabs.harvard.edu/abs/2019MNRAS.484..294D} {484, 294}

\bibitem[\protect\citeauthoryear{{Gaia Collab.} et~al.,}{{Gaia Collab.}
  et~al.}{2016}]{gaia+16}
{Gaia Collab.} et~al., 2016, \mn@doi [\aap] {10.1051/0004-6361/201629272},
  \href {http://adsabs.harvard.edu/abs/2016A%26A...595A...1G} {595, A1}

\bibitem[\protect\citeauthoryear{{Gaia Collab.} et~al.,}{{Gaia Collab.}
  et~al.}{2018}]{gaia+18}
{Gaia Collab.} et~al., 2018, \mn@doi [\aap] {10.1051/0004-6361/201833051},
  \href {https://ui.adsabs.harvard.edu/abs/2018A&A...616A...1G} {616, A1}

\bibitem[\protect\citeauthoryear{{Gaia Collab.} et~al.,}{{Gaia Collab.}
  et~al.}{2021}]{gaia+21}
{Gaia Collab.} et~al., 2021, \mn@doi [A\&A] {10.1051/0004-6361/202039657}, 649,
  A1

\bibitem[\protect\citeauthoryear{{Gibbons}, {Belokurov}  \& {Evans}}{{Gibbons}
  et~al.}{2017}]{gibbons+17}
{Gibbons} S.~L.~J.,  {Belokurov} V.,   {Evans} N.~W.,  2017, \mn@doi [\mnras]
  {10.1093/mnras/stw2328}, \href
  {https://ui.adsabs.harvard.edu/abs/2017MNRAS.464..794G} {464, 794}

\bibitem[\protect\citeauthoryear{{G{\'o}mez}, {Minchev}, {O'Shea}, {Beers},
  {Bullock}  \& {Purcell}}{{G{\'o}mez} et~al.}{2013}]{gomez+13}
{G{\'o}mez} F.~A.,  {Minchev} I.,  {O'Shea} B.~W.,  {Beers} T.~C.,  {Bullock}
  J.~S.,   {Purcell} C.~W.,  2013, \mn@doi [\mnras] {10.1093/mnras/sts327},
  \href {https://ui.adsabs.harvard.edu/abs/2013MNRAS.429..159G} {429, 159}

\bibitem[\protect\citeauthoryear{{Grand} et~al.,}{{Grand}
  et~al.}{2020}]{grand+20}
{Grand} R. J.~J.,  et~al., 2020, \mn@doi [\mnras] {10.1093/mnras/staa2057},
  \href {https://ui.adsabs.harvard.edu/abs/2020MNRAS.497.1603G} {497, 1603}

\bibitem[\protect\citeauthoryear{{Huang} et~al.,}{{Huang}
  et~al.}{2020}]{huang+20}
{Huang} Y.,  et~al., 2020, \mn@doi [\apjs] {10.3847/1538-4365/ab994f}, \href
  {https://ui.adsabs.harvard.edu/abs/2020ApJS..249...29H} {249, 29}

\bibitem[\protect\citeauthoryear{{Kazantzidis}, {Zentner}, {Kravtsov},
  {Bullock}  \& {Debattista}}{{Kazantzidis} et~al.}{2009}]{kazantzidis+09}
{Kazantzidis} S.,  {Zentner} A.~R.,  {Kravtsov} A.~V.,  {Bullock} J.~S.,
  {Debattista} V.~P.,  2009, \mn@doi [\apj] {10.1088/0004-637X/700/2/1896},
  \href {https://ui.adsabs.harvard.edu/abs/2009ApJ...700.1896K} {700, 1896}

\bibitem[\protect\citeauthoryear{{Kordopatis} et~al.,}{{Kordopatis}
  et~al.}{2023}]{kordopatis+23}
{Kordopatis} G.,  et~al., 2023, \mn@doi [\aap] {10.1051/0004-6361/202244283},
  \href {https://ui.adsabs.harvard.edu/abs/2023A&A...669A.104K} {669, A104}

\bibitem[\protect\citeauthoryear{Laporte, Johnston, Gómez, Garavito-Camargo
  \& Besla}{Laporte et~al.}{2018}]{laporte+18}
Laporte C. F.~P.,  Johnston K.~V.,  Gómez F.~A.,  Garavito-Camargo N.,   Besla
  G.,  2018, \mn@doi [Monthly Notices of the Royal Astronomical Society]
  {10.1093/mnras/sty1574}, 481, 286

\bibitem[\protect\citeauthoryear{{Laporte}, {Minchev}, {Johnston}  \&
  {G{\'o}mez}}{{Laporte} et~al.}{2019}]{laporte+19}
{Laporte} C. F.~P.,  {Minchev} I.,  {Johnston} K.~V.,   {G{\'o}mez} F.~A.,
  2019, \mn@doi [\mnras] {10.1093/mnras/stz583}, \href
  {https://ui.adsabs.harvard.edu/abs/2019MNRAS.485.3134L} {485, 3134}

\bibitem[\protect\citeauthoryear{{Lindegren} et~al.,}{{Lindegren}
  et~al.}{2018}]{lindegren+18}
{Lindegren} L.,  et~al., 2018, \mn@doi [\aap] {10.1051/0004-6361/201832727},
  \href {https://ui.adsabs.harvard.edu/abs/2018A&A...616A...2L} {616, A2}

\bibitem[\protect\citeauthoryear{{Mackereth} et~al.,}{{Mackereth}
  et~al.}{2017}]{mackereth+17}
{Mackereth} J.~T.,  et~al., 2017, \mn@doi [\mnras] {10.1093/mnras/stx1774},
  \href {https://ui.adsabs.harvard.edu/abs/2017MNRAS.471.3057M} {471, 3057}

\bibitem[\protect\citeauthoryear{{Martig}, {Minchev}  \& {Flynn}}{{Martig}
  et~al.}{2014a}]{martig+14a}
{Martig} M.,  {Minchev} I.,   {Flynn} C.,  2014a, \mn@doi [\mnras]
  {10.1093/mnras/stu1003}, \href
  {https://ui.adsabs.harvard.edu/abs/2014MNRAS.442.2474M} {442, 2474}

\bibitem[\protect\citeauthoryear{{Martig}, {Minchev}  \& {Flynn}}{{Martig}
  et~al.}{2014b}]{martig+14b}
{Martig} M.,  {Minchev} I.,   {Flynn} C.,  2014b, \mn@doi [\mnras]
  {10.1093/mnras/stu1322}, \href
  {https://ui.adsabs.harvard.edu/abs/2014MNRAS.443.2452M} {443, 2452}

\bibitem[\protect\citeauthoryear{{McMillan}}{{McMillan}}{2017}]{mcmillan17}
{McMillan} P.~J.,  2017, \mn@doi [\mnras] {10.1093/mnras/stw2759}, \href
  {https://ui.adsabs.harvard.edu/abs/2017MNRAS.465...76M} {465, 76}

\bibitem[\protect\citeauthoryear{{Minchev}, {Famaey}, {Quillen}, {Di Matteo},
  {Combes}, {Vlaji{\'c}}, {Erwin}  \& {Bland-Hawthorn}}{{Minchev}
  et~al.}{2012a}]{minchev+12a}
{Minchev} I.,  {Famaey} B.,  {Quillen} A.~C.,  {Di Matteo} P.,  {Combes} F.,
  {Vlaji{\'c}} M.,  {Erwin} P.,   {Bland-Hawthorn} J.,  2012a, \mn@doi [\aap]
  {10.1051/0004-6361/201219198}, \href
  {https://ui.adsabs.harvard.edu/abs/2012A&A...548A.126M} {548, A126}

\bibitem[\protect\citeauthoryear{{Minchev}, {Famaey}, {Quillen}, {Dehnen},
  {Martig}  \& {Siebert}}{{Minchev} et~al.}{2012b}]{minchev+12b}
{Minchev} I.,  {Famaey} B.,  {Quillen} A.~C.,  {Dehnen} W.,  {Martig} M.,
  {Siebert} A.,  2012b, \mn@doi [\aap] {10.1051/0004-6361/201219714}, \href
  {https://ui.adsabs.harvard.edu/abs/2012A&A...548A.127M} {548, A127}

\bibitem[\protect\citeauthoryear{{Mints} \& {Alexey}}{{Mints} \&
  {Alexey}}{2018}]{mints18}
{Mints} {Alexey} 2018, arXiv e-prints, \href
  {https://ui.adsabs.harvard.edu/abs/2018arXiv180501640M} {p. arXiv:1805.01640}

\bibitem[\protect\citeauthoryear{{Mucciarelli}, {Bellazzini}, {Ibata},
  {Romano}, {Chapman}  \& {Monaco}}{{Mucciarelli}
  et~al.}{2017}]{mucciarelli+17}
{Mucciarelli} A.,  {Bellazzini} M.,  {Ibata} R.,  {Romano} D.,  {Chapman}
  S.~C.,   {Monaco} L.,  2017, \mn@doi [\aap] {10.1051/0004-6361/201730707},
  \href {https://ui.adsabs.harvard.edu/abs/2017A&A...605A..46M} {605, A46}

\bibitem[\protect\citeauthoryear{Niederste-Ostholt, Belokurov, Evans  \&
  Peñarrubia}{Niederste-Ostholt et~al.}{2010}]{niederste_ostholt+10}
Niederste-Ostholt M.,  Belokurov V.,  Evans N.~W.,   Peñarrubia J.,  2010,
  \mn@doi [The Astrophysical Journal] {10.1088/0004-637X/712/1/516}, 712, 516

\bibitem[\protect\citeauthoryear{{Pietrzy{\'n}ski} et~al.,}{{Pietrzy{\'n}ski}
  et~al.}{2013}]{pietr+13}
{Pietrzy{\'n}ski} G.,  et~al., 2013, \mn@doi [\nat] {10.1038/nature11878},
  \href {https://ui.adsabs.harvard.edu/abs/2013Natur.495...76P} {495, 76}

\bibitem[\protect\citeauthoryear{{Piffl} et~al.,}{{Piffl}
  et~al.}{2014}]{piffl+14}
{Piffl} T.,  et~al., 2014, \mn@doi [\mnras] {10.1093/mnras/stu1948}, 445, 3133

\bibitem[\protect\citeauthoryear{{Queiroz} et~al.,}{{Queiroz}
  et~al.}{2018}]{quieroz+18}
{Queiroz} A.~B.~A.,  et~al., 2018, \mn@doi [\mnras] {10.1093/mnras/sty330},
  \href {https://ui.adsabs.harvard.edu/abs/2018MNRAS.476.2556Q} {476, 2556}

\bibitem[\protect\citeauthoryear{{Renaud}, {Agertz}, {Read}, {Ryde},
  {Andersson}, {Bensby}, {Rey}  \& {Feuillet}}{{Renaud}
  et~al.}{2021a}]{renaud+21a}
{Renaud} F.,  {Agertz} O.,  {Read} J.~I.,  {Ryde} N.,  {Andersson} E.~P.,
  {Bensby} T.,  {Rey} M.~P.,   {Feuillet} D.~K.,  2021a, \mn@doi [\mnras]
  {10.1093/mnras/stab250}, \href
  {https://ui.adsabs.harvard.edu/abs/2021MNRAS.503.5846R} {503, 5846}

\bibitem[\protect\citeauthoryear{{Renaud}, {Agertz}, {Andersson}, {Read},
  {Ryde}, {Bensby}, {Rey}  \& {Feuillet}}{{Renaud} et~al.}{2021b}]{renaud+21b}
{Renaud} F.,  {Agertz} O.,  {Andersson} E.~P.,  {Read} J.~I.,  {Ryde} N.,
  {Bensby} T.,  {Rey} M.~P.,   {Feuillet} D.~K.,  2021b, \mn@doi [\mnras]
  {10.1093/mnras/stab543}, \href
  {https://ui.adsabs.harvard.edu/abs/2021MNRAS.503.5868R} {503, 5868}

\bibitem[\protect\citeauthoryear{{Sanders} \& {Das}}{{Sanders} \&
  {Das}}{2018}]{sanders+18}
{Sanders} J.~L.,  {Das} P.,  2018, \mn@doi [\mnras] {10.1093/mnras/sty2490},
  \href {https://ui.adsabs.harvard.edu/abs/2018MNRAS.481.4093S} {481, 4093}

\bibitem[\protect\citeauthoryear{{Santistevan}, {Wetzel}, {Tollerud},
  {Sanderson}  \& {Samuel}}{{Santistevan} et~al.}{2023}]{santistevan+22}
{Santistevan} I.~B.,  {Wetzel} A.,  {Tollerud} E.,  {Sanderson} R.~E.,
  {Samuel} J.,  2023, \mn@doi [\mnras] {10.1093/mnras/stac3100}, \href
  {https://ui.adsabs.harvard.edu/abs/2023MNRAS.518.1427S} {518, 1427}

\bibitem[\protect\citeauthoryear{Santos-Santos, Sales, Fattahi  \&
  Navarro}{Santos-Santos et~al.}{2022}]{santos+22}
Santos-Santos I. M.~E.,  Sales L.~V.,  Fattahi A.,   Navarro J.~F.,  2022,
  \mn@doi [Monthly Notices of the Royal Astronomical Society]
  {10.1093/mnras/stac2057}, 515, 3685

\bibitem[\protect\citeauthoryear{{Sharma} et~al.,}{{Sharma}
  et~al.}{2021}]{sharma+21}
{Sharma} S.,  et~al., 2021, \mn@doi [\mnras] {10.1093/mnras/stab1086}, \href
  {https://ui.adsabs.harvard.edu/abs/2021MNRAS.506.1761S} {506, 1761}

\bibitem[\protect\citeauthoryear{{Thomas} et~al.,}{{Thomas}
  et~al.}{2019}]{thomas+19}
{Thomas} G.~F.,  et~al., 2019, \mn@doi [\mnras] {10.1093/mnras/sty3334}, \href
  {https://ui.adsabs.harvard.edu/abs/2019MNRAS.483.3119T} {483, 3119}

\bibitem[\protect\citeauthoryear{{Vasiliev}}{{Vasiliev}}{2019}]{vasiliev+19}
{Vasiliev} E.,  2019, \mn@doi [\mnras] {10.1093/mnras/sty2672}, \href
  {https://ui.adsabs.harvard.edu/abs/2019MNRAS.482.1525V} {482, 1525}

\bibitem[\protect\citeauthoryear{{Vasiliev} \& {Belokurov}}{{Vasiliev} \&
  {Belokurov}}{2020}]{vasiliev+20}
{Vasiliev} E.,  {Belokurov} V.,  2020, \mn@doi [\mnras]
  {10.1093/mnras/staa2114}, \href
  {https://ui.adsabs.harvard.edu/abs/2020MNRAS.497.4162V} {497, 4162}

\bibitem[\protect\citeauthoryear{{Velazquez} \& {White}}{{Velazquez} \&
  {White}}{1999}]{velazquez+99}
{Velazquez} H.,  {White} S. D.~M.,  1999, \mn@doi [\mnras]
  {10.1046/j.1365-8711.1999.02354.x}, \href
  {https://ui.adsabs.harvard.edu/abs/1999MNRAS.304..254V} {304, 254}

\bibitem[\protect\citeauthoryear{{Vera-Ciro} \& {D'Onghia}}{{Vera-Ciro} \&
  {D'Onghia}}{2016}]{veraciro+16}
{Vera-Ciro} C.,  {D'Onghia} E.,  2016, \mn@doi [\apj]
  {10.3847/0004-637X/824/1/39}, \href
  {https://ui.adsabs.harvard.edu/abs/2016ApJ...824...39V} {824, 39}

\bibitem[\protect\citeauthoryear{{Vera-Ciro}, {D'Onghia}, {Navarro}  \&
  {Abadi}}{{Vera-Ciro} et~al.}{2014}]{veraciro+14}
{Vera-Ciro} C.,  {D'Onghia} E.,  {Navarro} J.,   {Abadi} M.,  2014, \mn@doi
  [\apj] {10.1088/0004-637X/794/2/173}, \href
  {https://ui.adsabs.harvard.edu/abs/2014ApJ...794..173V} {794, 173}

\bibitem[\protect\citeauthoryear{{Yu} et~al.,}{{Yu} et~al.}{2021}]{zheng+21}
{Yu} Z.,  et~al., 2021, \mn@doi [\apj] {10.3847/1538-4357/abf098}, \href
  {https://ui.adsabs.harvard.edu/abs/2021ApJ...912..106Y} {912, 106}

\makeatother
\end{thebibliography}

\appendix

\label{lastpage}

\end{document}